\documentclass[twocolumn,letterpaper,aps,prc,longbibliography,superscriptaddress,nofootinbib,floatfix]{revtex4-1}

\usepackage{graphicx}	
\usepackage{ltxtable}
\usepackage{amssymb}
\usepackage{amsmath}
\usepackage{verbatim}
\usepackage{xspace}	

\newcommand{\pt}{\mbox{$p_T$}\xspace}

\newcommand{\rab}{\mbox{$R_{AB}$}\xspace}

\newcommand{\Ncoll}{\mbox{$N_{\mathrm coll}$}\xspace}

\newcommand{\ncoll}{\mbox{$N_{\mathrm coll}$}\xspace}
\newcommand{\meanncoll}{\mbox{$\langle N_{\mathrm coll} \rangle$}\xspace}

\newcommand{\meanptsq}{\mbox{$\langle p_T^2 \rangle$}\xspace}

\newcommand{\sqsntwo}{\mbox{$\sqrt{s_{_{NN}}}=200~\mathrm{GeV}$}\xspace}

\newcommand{\sqstwo}{\mbox{$\sqrt{s}=200~\mathrm{GeV}$}\xspace}
\newcommand{\pp}{\mbox{$p$$+$$p$}\xspace}
\newcommand{\pA}{\mbox{$p$$+$$A$}\xspace}
\newcommand{\dau}{\mbox{$d$$+$Au}\xspace}

\newcommand{\pau}{\mbox{$p$$+$Au}\xspace}
\newcommand{\pal}{\mbox{$p$$+$Al}\xspace}
\newcommand{\ppb}{\mbox{$p$$+$Pb}\xspace}
\newcommand{\heau}{\mbox{$^{3}$He$+$Au}\xspace}

\newcommand{\mumu}{\mbox{$\mu^{+}\mu^{-}$}\xspace}

\newcommand{\jpsi}{\mbox{$J/\psi$}\xspace}

\newcommand{\pythia}{\mbox{\textsc{pythia8}}\xspace}
\newcommand{\geant}{\mbox{\textsc{geant4}}\xspace}

\begin{document}

\title{Measurement of $J/\psi$ at forward and backward rapidity in $p$+$p$, 
$p$$+$Al, $p$$+$Au, and $^3$He+Au collisions at $\sqrt{s_{_{NN}}}=200~{\rm GeV}$}

\newcommand{\abilene}{Abilene Christian University, Abilene, Texas 79699, USA}
\newcommand{\augie}{Department of Physics, Augustana University, Sioux Falls, South Dakota 57197, USA}
\newcommand{\banaras}{Department of Physics, Banaras Hindu University, Varanasi 221005, India}
\newcommand{\barc}{Bhabha Atomic Research Centre, Bombay 400 085, India}
\newcommand{\baruch}{Baruch College, City University of New York, New York, New York, 10010 USA}
\newcommand{\bnlcoll}{Collider-Accelerator Department, Brookhaven National Laboratory, Upton, New York 11973-5000, USA}
\newcommand{\bnlphys}{Physics Department, Brookhaven National Laboratory, Upton, New York 11973-5000, USA}
\newcommand{\caucr}{University of California-Riverside, Riverside, California 92521, USA}
\newcommand{\charlesczech}{Charles University, Ovocn\'{y} trh 5, Praha 1, 116 36, Prague, Czech Republic}
\newcommand{\ciae}{Science and Technology on Nuclear Data Laboratory, China Institute of Atomic Energy, Beijing 102413, People's Republic of China}
\newcommand{\cns}{Center for Nuclear Study, Graduate School of Science, University of Tokyo, 7-3-1 Hongo, Bunkyo, Tokyo 113-0033, Japan}
\newcommand{\colorado}{University of Colorado, Boulder, Colorado 80309, USA}
\newcommand{\columbia}{Columbia University, New York, New York 10027 and Nevis Laboratories, Irvington, New York 10533, USA}
\newcommand{\czechtech}{Czech Technical University, Zikova 4, 166 36 Prague 6, Czech Republic}
\newcommand{\debrecen}{Debrecen University, H-4010 Debrecen, Egyetem t{\'e}r 1, Hungary}
\newcommand{\elte}{ELTE, E{\"o}tv{\"o}s Lor{\'a}nd University, H-1117 Budapest, P{\'a}zm{\'a}ny P.~s.~1/A, Hungary}
\newcommand{\eszterhazy}{Eszterh\'azy K\'aroly University, K\'aroly R\'obert Campus, H-3200 Gy\"ongy\"os, M\'atrai \'ut 36, Hungary}
\newcommand{\ewha}{Ewha Womans University, Seoul 120-750, Korea}
\newcommand{\famu}{Florida A\&M University, Tallahassee, FL 32307, USA}
\newcommand{\fsu}{Florida State University, Tallahassee, Florida 32306, USA}
\newcommand{\gsu}{Georgia State University, Atlanta, Georgia 30303, USA}
\newcommand{\hiroshima}{Hiroshima University, Kagamiyama, Higashi-Hiroshima 739-8526, Japan}
\newcommand{\howard}{Department of Physics and Astronomy, Howard University, Washington, DC 20059, USA}
\newcommand{\ihepprot}{IHEP Protvino, State Research Center of Russian Federation, Institute for High Energy Physics, Protvino, 142281, Russia}
\newcommand{\illuiuc}{University of Illinois at Urbana-Champaign, Urbana, Illinois 61801, USA}
\newcommand{\inrras}{Institute for Nuclear Research of the Russian Academy of Sciences, prospekt 60-letiya Oktyabrya 7a, Moscow 117312, Russia}
\newcommand{\instpasczech}{Institute of Physics, Academy of Sciences of the Czech Republic, Na Slovance 2, 182 21 Prague 8, Czech Republic}
\newcommand{\isu}{Iowa State University, Ames, Iowa 50011, USA}
\newcommand{\jaea}{Advanced Science Research Center, Japan Atomic Energy Agency, 2-4 Shirakata Shirane, Tokai-mura, Naka-gun, Ibaraki-ken 319-1195, Japan}
\newcommand{\jeonbuk}{Jeonbuk National University, Jeonju, 54896, Korea}
\newcommand{\jyvaskyla}{Helsinki Institute of Physics and University of Jyv{\"a}skyl{\"a}, P.O.Box 35, FI-40014 Jyv{\"a}skyl{\"a}, Finland}
\newcommand{\kek}{KEK, High Energy Accelerator Research Organization, Tsukuba, Ibaraki 305-0801, Japan}
\newcommand{\korea}{Korea University, Seoul, 02841}
\newcommand{\kurchatov}{National Research Center ``Kurchatov Institute", Moscow, 123098 Russia}
\newcommand{\kyoto}{Kyoto University, Kyoto 606-8502, Japan}
\newcommand{\lawllnl}{Lawrence Livermore National Laboratory, Livermore, California 94550, USA}
\newcommand{\losalamos}{Los Alamos National Laboratory, Los Alamos, New Mexico 87545, USA}
\newcommand{\lund}{Department of Physics, Lund University, Box 118, SE-221 00 Lund, Sweden}
\newcommand{\lyon}{IPNL, CNRS/IN2P3, Univ Lyon, Université Lyon 1, F-69622, Villeurbanne, France}
\newcommand{\maryland}{University of Maryland, College Park, Maryland 20742, USA}
\newcommand{\mass}{Department of Physics, University of Massachusetts, Amherst, Massachusetts 01003-9337, USA}
\newcommand{\michigan}{Department of Physics, University of Michigan, Ann Arbor, Michigan 48109-1040, USA}
\newcommand{\muhlenberg}{Muhlenberg College, Allentown, Pennsylvania 18104-5586, USA}
\newcommand{\nara}{Nara Women's University, Kita-uoya Nishi-machi Nara 630-8506, Japan}
\newcommand{\natmephi}{National Research Nuclear University, MEPhI, Moscow Engineering Physics Institute, Moscow, 115409, Russia}
\newcommand{\newmex}{University of New Mexico, Albuquerque, New Mexico 87131, USA}
\newcommand{\nmsu}{New Mexico State University, Las Cruces, New Mexico 88003, USA}
\newcommand{\northcg}{Physics and Astronomy Department, University of North Carolina at Greensboro, Greensboro, North Carolina 27412, USA}
\newcommand{\ohio}{Department of Physics and Astronomy, Ohio University, Athens, Ohio 45701, USA}
\newcommand{\ornl}{Oak Ridge National Laboratory, Oak Ridge, Tennessee 37831, USA}
\newcommand{\orsay}{IPN-Orsay, Univ.~Paris-Sud, CNRS/IN2P3, Universit\'e Paris-Saclay, BP1, F-91406, Orsay, France}
\newcommand{\peking}{Peking University, Beijing 100871, People's Republic of China}
\newcommand{\pnpi}{PNPI, Petersburg Nuclear Physics Institute, Gatchina, Leningrad region, 188300, Russia}
\newcommand{\pusan}{Pusan National University, Busan, 46241, South Korea}
\newcommand{\riken}{RIKEN Nishina Center for Accelerator-Based Science, Wako, Saitama 351-0198, Japan}
\newcommand{\rikjrbrc}{RIKEN BNL Research Center, Brookhaven National Laboratory, Upton, New York 11973-5000, USA}
\newcommand{\rikkyo}{Physics Department, Rikkyo University, 3-34-1 Nishi-Ikebukuro, Toshima, Tokyo 171-8501, Japan}
\newcommand{\saispbstu}{Saint Petersburg State Polytechnic University, St.~Petersburg, 195251 Russia}
\newcommand{\seoulnat}{Department of Physics and Astronomy, Seoul National University, Seoul 151-742, Korea}
\newcommand{\stonybrkc}{Chemistry Department, Stony Brook University, SUNY, Stony Brook, New York 11794-3400, USA}
\newcommand{\stonycrkp}{Department of Physics and Astronomy, Stony Brook University, SUNY, Stony Brook, New York 11794-3800, USA}
\newcommand{\tenn}{University of Tennessee, Knoxville, Tennessee 37996, USA}
\newcommand{\titech}{Department of Physics, Tokyo Institute of Technology, Oh-okayama, Meguro, Tokyo 152-8551, Japan}
\newcommand{\tsukuba}{Tomonaga Center for the History of the Universe, University of Tsukuba, Tsukuba, Ibaraki 305, Japan}
\newcommand{\vandy}{Vanderbilt University, Nashville, Tennessee 37235, USA}
\newcommand{\weizmann}{Weizmann Institute, Rehovot 76100, Israel}
\newcommand{\wigner}{Institute for Particle and Nuclear Physics, Wigner Research Centre for Physics, Hungarian Academy of Sciences (Wigner RCP, RMKI) H-1525 Budapest 114, POBox 49, Budapest, Hungary}
\newcommand{\yonsei}{Yonsei University, IPAP, Seoul 120-749, Korea}
\newcommand{\zagreb}{Department of Physics, Faculty of Science, University of Zagreb, Bijeni\v{c}ka c.~32 HR-10002 Zagreb, Croatia}
\affiliation{\abilene}
\affiliation{\augie}
\affiliation{\banaras}
\affiliation{\barc}
\affiliation{\baruch}
\affiliation{\bnlcoll}
\affiliation{\bnlphys}
\affiliation{\caucr}
\affiliation{\charlesczech}
\affiliation{\ciae}
\affiliation{\cns}
\affiliation{\colorado}
\affiliation{\columbia}
\affiliation{\czechtech}
\affiliation{\debrecen}
\affiliation{\elte}
\affiliation{\eszterhazy}
\affiliation{\ewha}
\affiliation{\famu}
\affiliation{\fsu}
\affiliation{\gsu}
\affiliation{\hiroshima}
\affiliation{\howard}
\affiliation{\ihepprot}
\affiliation{\illuiuc}
\affiliation{\inrras}
\affiliation{\instpasczech}
\affiliation{\isu}
\affiliation{\jaea}
\affiliation{\jeonbuk}
\affiliation{\jyvaskyla}
\affiliation{\kek}
\affiliation{\korea}
\affiliation{\kurchatov}
\affiliation{\kyoto}
\affiliation{\lawllnl}
\affiliation{\losalamos}
\affiliation{\lund}
\affiliation{\lyon}
\affiliation{\maryland}
\affiliation{\mass}
\affiliation{\michigan}
\affiliation{\muhlenberg}
\affiliation{\nara}
\affiliation{\natmephi}
\affiliation{\newmex}
\affiliation{\nmsu}
\affiliation{\northcg}
\affiliation{\ohio}
\affiliation{\ornl}
\affiliation{\orsay}
\affiliation{\peking}
\affiliation{\pnpi}
\affiliation{\pusan}
\affiliation{\riken}
\affiliation{\rikjrbrc}
\affiliation{\rikkyo}
\affiliation{\saispbstu}
\affiliation{\seoulnat}
\affiliation{\stonybrkc}
\affiliation{\stonycrkp}
\affiliation{\tenn}
\affiliation{\titech}
\affiliation{\tsukuba}
\affiliation{\vandy}
\affiliation{\weizmann}
\affiliation{\wigner}
\affiliation{\yonsei}
\affiliation{\zagreb}
\author{U.~Acharya} \affiliation{\gsu} 
\author{A.~Adare} \affiliation{\colorado} 
\author{C.~Aidala} \affiliation{\michigan} 
\author{N.N.~Ajitanand} \altaffiliation{Deceased} \affiliation{\stonybrkc} 
\author{Y.~Akiba} \email[PHENIX Spokesperson: ]{akiba@rcf.rhic.bnl.gov} \affiliation{\riken} \affiliation{\rikjrbrc} 
\author{M.~Alfred} \affiliation{\howard} 
\author{V.~Andrieux} \affiliation{\michigan} 
\author{N.~Apadula} \affiliation{\isu} \affiliation{\stonycrkp} 
\author{H.~Asano} \affiliation{\kyoto} \affiliation{\riken} 
\author{B.~Azmoun} \affiliation{\bnlphys} 
% ...
\collaboration{PHENIX Collaboration} \noaffiliation

\date{\today}


\begin{abstract}


Charmonium is a valuable probe in heavy-ion collisions to study the
properties of the quark gluon plasma, and is also an interesting probe
in small collision systems to study cold nuclear matter effects, which
are also present in large collision systems. With the recent
observations of collective behavior of produced particles in small
system collisions, measurements of the modification of charmonium in
small systems have become increasingly relevant. We present the results
of $J/\psi$ measurements at forward and backward rapidity in various
small collision systems, $p$$+$$p$, $p$$+$Al, $p$$+$Au and $^3$He$+$Au,
at $\sqrt{s_{_{NN}}}$=200 GeV. The results are presented in the form of the
observable $R_{AB}$, the nuclear modification factor, a measure of the
ratio of the $J/\psi$ invariant yield compared to the scaled yield in
$p$$+$$p$ collisions. We examine the rapidity, transverse momentum, and
collision centrality dependence of nuclear effects on $J/\psi$ production
with different projectile sizes $p$ and $^3$He, and different target
sizes Al and Au. The modification is found to be strongly dependent on
the target size, but to be very similar for $p$$+$Au and $^{3}$He$+$Au.
However, for 0\%--20\% central collisions at backward rapidity, the
modification factor for $^{3}$He$+$Au is found to be smaller than that for
$p$$+$Au, with a mean fit to the ratio of
$0.89\pm0.03$(stat)${\pm}0.08$(syst), possibly indicating final state
effects due to the larger projectile size.

\end{abstract}

\maketitle

\section{Introduction}
\label{sec:introduction}

The cross section for production of charmonium in proton collisions with 
heavy nuclei is strongly modified relative to that in \pp collisions. 
The effects that cause this modification are often referred to as cold 
nuclear matter (CNM) effects because of the long-standing presumption 
that the energy density and temperature produced in the collision of a 
single proton with a nucleus were not sufficient to form a deconfined 
quark-gluon plasma, as produced in ultra-relativistic heavy ion 
collisions at the Relativistic Heavy Ion Collider (RHIC) and the Large 
Hadron Collider (LHC).  A major motivation for this work is to study CNM 
effects that can modify charm production in \pA collisions, which 
include modification of the nuclear-parton-distribution
functions (nPDFs) in a nucleus~\cite{Eskola:2016oht,Kovarik:2015cma}, 
initial state parton energy loss~\cite{Vitev:2007ve}, breakup of the 
forming charmonium in collisions with target 
nucleons~\cite{McGlinchey:2012bp,Arleo:1999af}, coherent gluon 
saturation~\cite{Kharzeev:2005zr,Fujii:2006ab}, and transverse momentum 
broadening~\cite{Cronin:1974zm}.  These mechanisms are generally expected 
to act in the early stages of the collision, and effect either the 
production rates of charm quarks or their propagation through the 
nucleus.  All of these processes are strongly (and differently) 
dependent on the rapidity and transverse momentum of the produced 
charmonium, and the collision energy. They are therefore best studied 
using \pA data covering the broadest possible range of collision energy, 
rapidity and transverse momentum.

At RHIC, \pp, \dau, \pau, \heau and 
\pal collisions have been studied at \sqsntwo. The PHENIX experiment 
published data on \jpsi production in \dau collisions over the rapidity 
intervals $1.2<|y|<2.2$ and $|y|<0.35$~\cite{Adare:2010fn, 
Adare:2012qf}. PHENIX also reported measurements of the $\psi(2S)$ in 
small collision systems, first with nuclear modification in $d$$+$Au 
collisions ($|y|<0.35$)~\cite{Adare:2013ezl}, followed by measurements 
of the ratio of $\psi(2S)$ to $J/\psi$ in \pal, \pau and \heau 
collisions at \sqsntwo ($1.2<|y|<2.2$)~\cite{Adare:2016psx}. The STAR 
collaboration has reported \jpsi nuclear modification data for \dau 
collisions ($|y|<1$)~\cite{Adamczyk:2016dhc}.

At the LHC, nuclear effects in \ppb collisions 
have been studied at $\sqrt{s_{_{NN}}}=5.02$ TeV. The ALICE 
collaboration has reported data for 
\jpsi~\cite{Adam:2015jsa,Abelev:2013yxa} and 
$\psi(2S)$~\cite{Abelev:2014zpa, Adam:2016ohd} ($-4.46<y<-2.96$ and 
$2.03<y<3.53$). The LHCb collaboration has reported 
\jpsi~\cite{Aaij:2013zxa} and $\psi(2S)$ data~\cite{Aaij:2016eyl} 
($-5.0<y<-2.5$ and $1.5<y<4.0$). The CMS collaboration has reported 
\jpsi~\cite{Sirunyan:2017mzd} and $\psi(2S)$~\cite{Sirunyan:2018pse} 
data ($-2.4<y<1.93$ and $p_T>4~\mathrm{GeV}/c$). The ATLAS collaboration 
has reported \jpsi~\cite{Aad:2015ddl} and 
charmonium~\cite{Aaboud:2017cif} data ($|y|<2$ and $p_T>8$~GeV/$c$).  
These measurements show a significant energy, rapidity and \pt 
dependence of the modification of charmonia production compared to the 
scaled \pp results.

The assumption that effects due to soft particles produced in the collision 
are not important in $p$ or $d$$+$$A$ collision at colliders was called into 
question by the observation of strong suppression of the $\psi(2S)$ relative 
to the \jpsi in central \dau collisions~\cite{Adare:2013ezl}, and then in 
\ppb collisions~\cite{Abelev:2014zpa}. Because CNM effects on the production 
of charm quarks and their transport through the nucleus are expected to 
affect both states similarly, they do not appear to be able to explain this 
observation. However, it can be reproduced by the co-mover break up 
model~\cite{Ferreiro:2014bia}, where charmonium is dissociated by 
interactions with produced particles in the final state, which naturally 
gives a larger suppression effect on the much more weakly bound $\psi(2S)$. 
The observation of flow-like behavior in \ppb collisions at LHC (see for 
example~\cite{Dusling:2015gta}) and later in \dau collisions at 
RHIC~\cite{Adare:2013piz, Adare:2014keg} suggested that a quark-gluon plasma 
of small size may be formed in high energy collisions of these light 
systems.  This led to the application of transport models to \ppb and \dau 
data, which were originally developed for charmonium production in heavy ion 
collisions~\cite{Du:2015wha, Beraudo:2015wsd}.  A plasma phase in these 
small collision systems gives different suppression between the charmonia 
states and allows a description of the data. In the case of most central 
midrapidity \dau collisions at \sqsntwo, additional suppression beyond CNM 
effects has been predicted of approximately 20\% for the \jpsi, and 55\% for 
the $\psi(2S)$~\cite{Du:2015wha}, in good agreement with the 
data~\cite{Adare:2010fn,Adare:2013ezl}.

In 2014 and 2015, RHIC provided collisions of \pal, \pau, and \heau for a 
systematic study of small systems. A comparison of flow data from \pau, 
\dau, and \heau with hydrodynamic models found that the data were all 
consistent with hydrodynamic flow in the most central 
collisions~\cite{PHENIX:2018lia,Koop:2015wea,Adare:2015ctn}. An obvious 
question is whether increased energy density provided by the $^3$He 
projectile in comparison to the proton produces any observable effect on 
charmonium modification in collisions with a Au target.

In this paper we present PHENIX measurements of inclusive \jpsi 
production in \pal, \pau, and \heau collisions at \sqsntwo.  The 
inclusive \jpsi cross section includes feed-down from $\psi(2S)$ and 
$\chi_c$ states, and a smaller contribution from B-meson decays.  The 
results are directly compared to $p$$+$$p$ collisions at the same center of 
mass energy by calculating the nuclear modification factor $R_{AB}$.  
The \jpsi data are presented as a function of \pt, rapidity, and 
centrality and are compared to theoretical models.

\section{Experimental Setup}
\label{sec:experiment}

\begin{figure*}[th]
\includegraphics[width=0.65\linewidth]{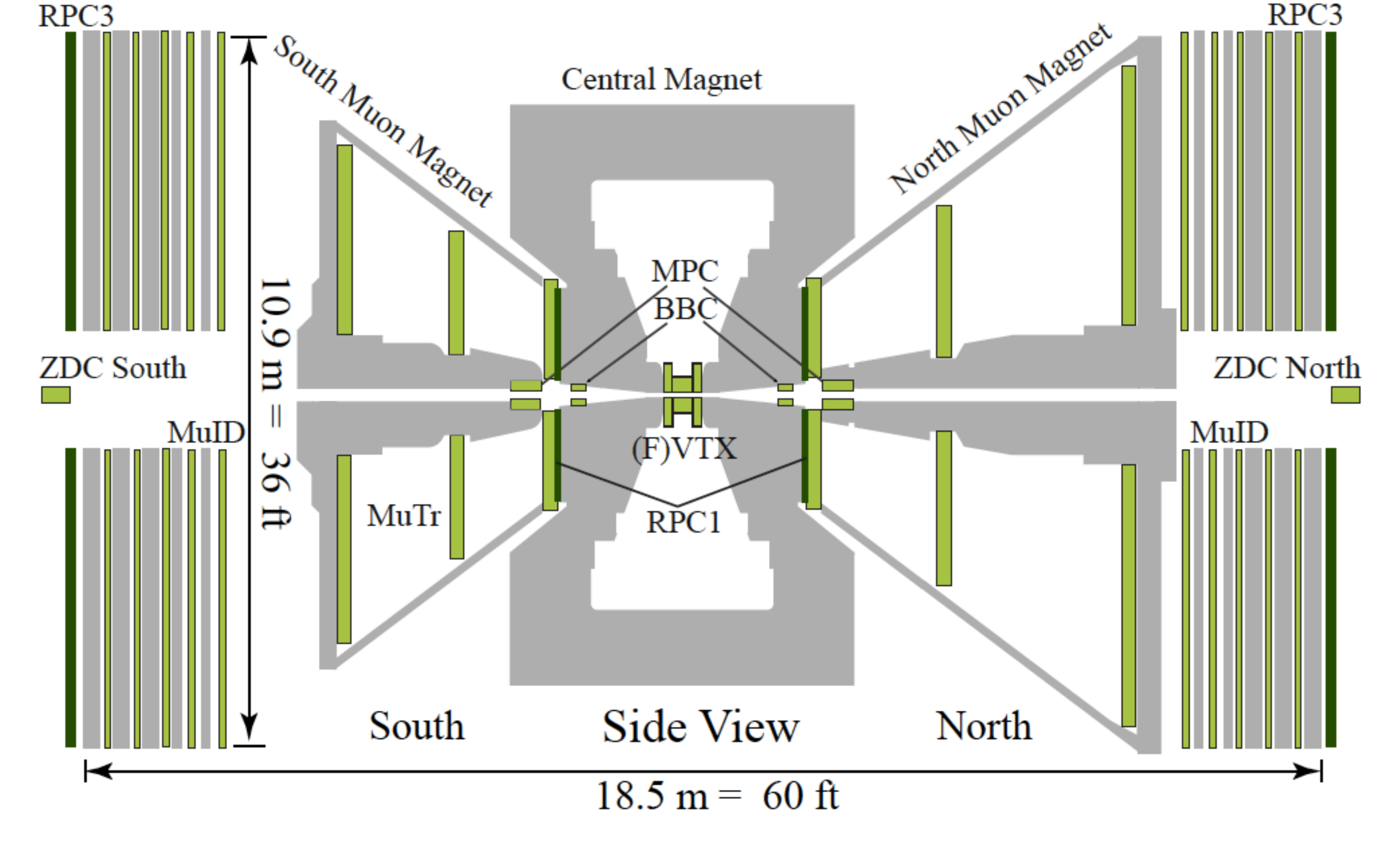}
\caption{\label{fig:PHENIX}
Side view of the PHENIX detector in 2014 and 2015.}
\end{figure*}

The PHENIX detector~\cite{Adcox:2003zm} comprises two central arm 
spectrometers at midrapidity and two muon arm spectrometers at forward 
and backward rapidity.  The detector configuration during the data 
taking in 2014 and 2015 is shown in Fig.~\ref{fig:PHENIX}. The data 
presented here are from $J/\psi \rightarrow \mu^+ \mu^-$ decays recorded 
with the muon arm spectrometers. The muon spectrometers have full 
azimuthal acceptance, covering $-2.2<\eta<-1.2$ (south arm) and 
$1.2<\eta<2.4$ (north arm), where the forward arm has a slightly larger 
acceptance than the backward arm. For dimuons, the analysis is 
restricted to $1.2<|y|<2.2$ in both arms. Each muon arm comprises a 
Forward Silicon Vertex Tracker (FVTX), followed by a hadron absorber and 
a muon spectrometer.

The FVTX~\cite{Aidala:2013vna} is a silicon detector designed to measure 
a precise collision vertex (also constrained by the Silicon Vertex 
Tracker (VTX) at midrapidity), and to provide precise tracking for 
charged particles entering the muon spectrometer before undergoing 
multiple scattering in the hadron absorber. The FVTX was not used in 
this inclusive \jpsi analysis, because the acceptance is reduced when 
requiring muon arm tracks that match tracks in the FVTX. Following the 
FVTX is the hadron absorber, composed of layers of copper, iron, and 
stainless steel, corresponding to 7.2 nuclear interaction lengths 
($\lambda_{I}$).  The absorber suppresses hadrons in front of the muon 
arm by a factor of approximately 1000, thus significantly reducing 
hadronic background for muon based measurements.

Each of the muon spectrometers is composed of a muon tracker (MuTr) 
embedded in a magnetic field followed by a muon identifier (MuID). Each 
MuTr comprises three stations of cathode strip chambers, inside a 
magnet with a radial field integral of $\int{{\rm B}\cdot dl}=0.72~{\rm 
T}\cdot{\rm m}$.  It provides a momentum measurement for charged 
particles. Each MuID is composed of five layers (referred to as gap 
0--4) of steel absorber (4.8 (5.4) $\lambda_{I}$ for south (north) arm) 
and two planes of Iarocci tubes.  This enables the separation of muons 
and hadrons based on their penetration depth at a given reconstructed 
momentum. The MuID in each arm is also used to trigger events containing 
two or more muon tracks per event, called a dimuon trigger, and each 
muon track is required to have at least one hit in either gap 3 or gap 
4. A more detailed discussion of the PHENIX muon arms can be found in 
Ref.~\cite{Akikawa:2003zs,Adachi:2013qha}.

The beam-beam counters (BBC) are used to determine the collision vertex 
position along the beam axis ($z_{\rm BBC}$) with a resolution of 
roughly 2~cm in \pp collisions. Each BBC comprises two arrays of 64 
quartz \v{C}erenkov detectors located at $z=\pm144~{\rm cm}$ from the 
nominal interaction point, and has an acceptance covering the full 
azimuth and $3.1<|y|<3.9$. They also provide a minimum bias (MB) 
trigger by requiring at least one hit in each BBC. The BBC trigger 
efficiency, determined from the Van der Meer scan 
technique~\cite{Drees:2003zza}, is 55\% $\pm$5\% for inelastic \pp 
events and 79\% $\pm$2\% for events with midrapidity particle 
production~\cite{Adler:2003pb,Adare:2013nff}. In \pal, \pau, and \heau 
collisions, charged particle multiplicity in the BBC in the Au/Al-going 
direction ($-3.9<y<-3.1$) is used to categorize the event centrality. 
The BBC trigger efficiency is 72\% $\pm$4\%, 84\% $\pm$3\%, and 88\% 
$\pm$4\% of inelastic \pal, \pau, and \heau collisions, respectively.

A Glauber model, combined with a simulation of the BBC response, is used 
to relate charged particle multiplicity in the BBC to parameters that 
characterize the collision centrality, as described 
in~\cite{Adare:2013nff}. The analysis produces the average number of 
nucleon-nucleon collisions in each centrality category. It also produces 
centrality dependent BBC bias correction factors which account for the 
correlation between BBC charge and the presence of a hard scattering in 
the event, and are applied as a multiplicative correction on invariant 
yields. Table~\ref{tab:centrality} shows the values of \meanncoll and 
BBC bias correction factor from this analysis.

\begin{table}[tbh]
\caption{\label{tab:centrality}
\meanncoll and BBC bias correction factors for different centrality bins 
of \pal, \pau and \heau collisions.
}
\begin{ruledtabular} \begin{tabular}{cccc}
 Collision system & Centrality & \meanncoll & Bias factor \\
\hline
 \pal & 0\%--20\%  & 3.4$\pm$0.3  & 0.81$\pm$0.01 \\
      & 20\%--40\% & 2.4$\pm$0.1  & 0.90$\pm$0.02 \\
      & 40\%--72\% & 1.7$\pm$0.1  & 1.04$\pm$0.04 \\  
      & 0\%--100\% & 2.1$\pm$0.1  & 0.80$\pm$0.02 \\
\\
 \pau & 0\%--5\%   & 9.7$\pm$0.6  & 0.86$\pm$0.01 \\
      & 5\%--10\%  & 8.4$\pm$0.6  & 0.90$\pm$0.01 \\
      & 10\%--20\% & 7.4$\pm$0.5  & 0.94$\pm$0.01 \\
      & 0\%--20\%  & 8.2$\pm$0.5  & 0.90$\pm$0.01 \\
      & 20\%--40\% & 6.1$\pm$0.4  & 0.98$\pm$0.01 \\
      & 40\%--60\% & 4.4$\pm$0.3  & 1.03$\pm$0.01 \\
      & 60\%--84\% & 2.6$\pm$0.2  & 1.00$\pm$0.06 \\
      & 0\%--100\% & 4.7$\pm$0.3  & 0.86$\pm$0.01 \\
\\
 \heau& 0\%--20\%  & 22.3$\pm$1.7 & 0.95$\pm$0.01 \\
      & 20\%--40\% & 14.8$\pm$1.1 & 0.95$\pm$0.01 \\
      & 40\%--88\% & 5.5$\pm$0.4  & 1.03$\pm$0.01 \\  
      & 0\%--100\% & 10.4$\pm$0.7 & 0.89$\pm$0.01 \\      
\end{tabular} \end{ruledtabular}
\end{table}

\section{Data analysis}

\subsection{Data set}

The data sets used in this analysis are \heau data collected in 2014, 
and \pp, \pal, and \pau data collected in 2015.  All data sets were 
recorded at a center of mass energy $\sqrt{s_{_{NN}}}$=200 GeV. The events 
considered here are triggered by the dimuon trigger and are required to 
have a vertex within $\pm 30~\mathrm{cm}$ of the center of the 
interaction region.  The corresponding integrated luminosity is 47 
pb$^{-1}$ for \pp, 590 nb$^{-1}$ for \pal, 138 nb$^{-1}$ for \pau, and 
18 nb$^{-1}$ for \heau collisions.

\subsection{\jpsi signal extraction}

\begin{figure}[htb]
\includegraphics[width=0.96\linewidth]{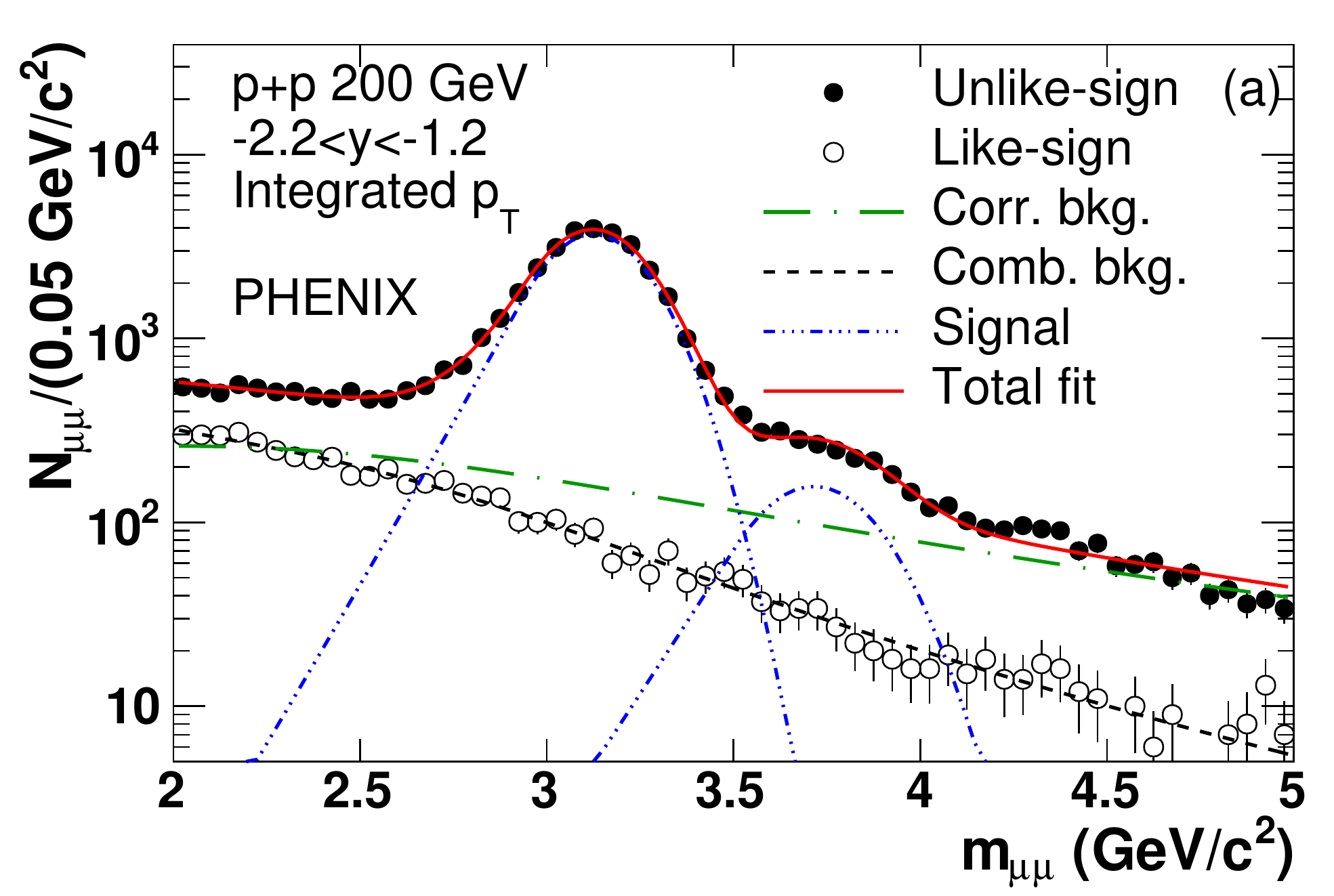}
\includegraphics[width=0.96\linewidth]{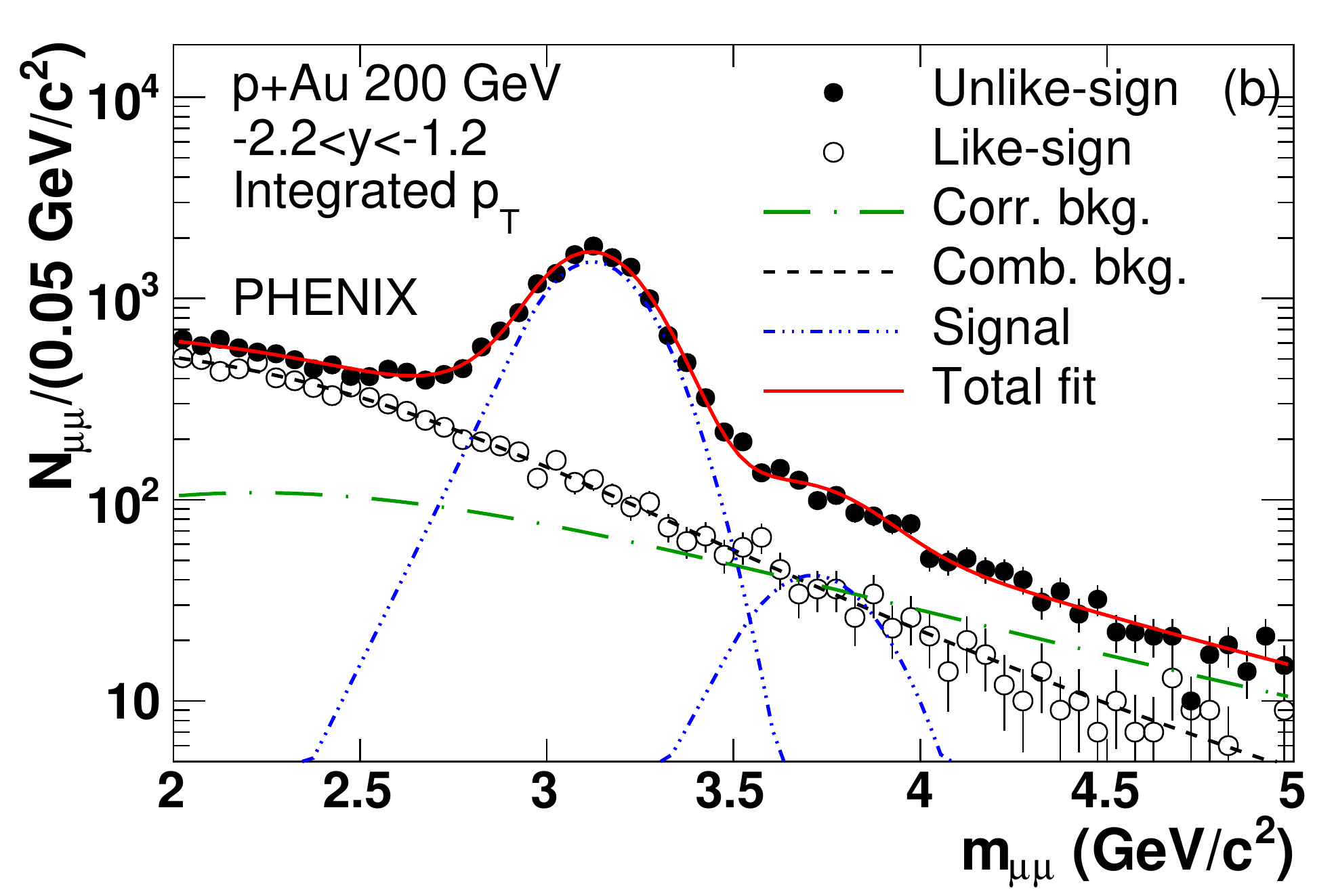}
\caption{\label{fig:fit}
Invariant mass distributions of unlike-sign and like-sign dimuons in \pp and integrated centrality of \pau collisions in the south muon arm. Fit results to extract 
the \jpsi signal are also presented.}
\end{figure}

Yields of \jpsi mesons were extracted from the invariant mass spectra 
constructed from combinations of unlike-sign tracks that are identified as 
muons (see Fig.~\ref{fig:fit}).  The mass spectra contain muon pairs from 
\jpsi decays, as well as significant contributions from combinations of real 
muons not from a \jpsi, as well as misidentified hadrons.  Details about the 
dimuon selection to reduce the background contributions are described 
in~\cite{Aidala:2018ajl,Adare:2011vq}.

The mass spectrum constructed from like-sign tracks was used to estimate 
the background due to random combinations of kinematically unrelated 
tracks. A modified Hagedorn function was used to represent the 
correlated background due to kinematically related tracks. For \jpsi 
signal extraction, Crystal-ball functions~\cite{Gaiser:1982yw} were used 
to describe the \jpsi and $\psi(2S)$ peaks, similar to the previous 
analysis in small collision systems~\cite{Adare:2016psx}:

\begin{multline}
f(m) = N  \cdot  {\rm exp} \left (-\frac{(m -\bar{m})^2}{2\sigma^2} \right ), \hspace {2mm} \small{{\rm for}\hspace{2mm} \frac{m-\bar{m}}{\sigma} > -\alpha }\\
f(m) = N  \cdot  A \cdot \left ( B -\frac{(m -\bar{m})^2}{\sigma} \right )^{-n}, \hspace {2mm} \small{{\rm for}\hspace{2mm} \frac{m-\bar{m}}{\sigma} \leq -\alpha},\\
A = \left ( \frac{n}{|\alpha|} \right )^n \cdot {\rm exp}\left ( 
-\frac{|\alpha|^2}{2} \right ), \hspace{2mm} B = \frac{n}{|\alpha|} - |\alpha|,
\end{multline}
where $\sigma$ and $\bar{m}$ are the width and mass centroid of the 
Gaussian component of the line shape and $\alpha$ and $n$ are parameters 
describing the tail.

The crystal-ball shape and tail parameters for the $\psi(2S)$ were fixed 
with respect to the \jpsi parameters, using the PDG database 
value~\cite{PDG} for the energy difference and a width broadening factor 
taken from simulations. In cases where the statistical precision of the 
data led to poor definition of the \jpsi signal shape, the mass and 
width of the \jpsi peak were fixed and a systematic uncertainty was 
assigned to the yield based on tests made with higher statistics cases.
The statistical uncertainties related to the extraction of the \jpsi 
yields were determined from a covariance matrix in the fitting 
procedure.

\subsection{Background estimation}
\label{sec:background}

The random combinatorial background in the unlike-sign mass spectrum was 
approximated by combining all like-sign tracks from the same events. 
There is a small correlated contribution to the like-sign pairs from 
jets and open bottom; however, compared to the other background sources, 
this is small.

The correlated background comprises unlike-sign muon pairs from charm, 
bottom, jets, and Drell-Yan. Because the correlated background cannot be 
estimated independently from the data, it must be fitted to the mass 
spectrum when the \jpsi yield is extracted. Fitting the correlated 
background effectively compensates for the small correlated component 
included in the like-sign estimation of the combinatorial background.

We describe the correlated background using a modified Hagedorn 
function~\cite{Adare:2010de, Aidala:2018ajl, Adare:2009qk}:

\begin{equation}
\label{eq:hagedorn}
   \frac{d^2N}{dm_{\mu\mu}d\pt} = \frac{p_{0}}{[\exp{(-p_{1}m_{\mu\mu} - 
p_{2}m_{\mu\mu}^2)} + {m_{\mu\mu}}/{p_{3}}]^{p_{4}}},
\end{equation}
where $m_{\mu \mu}$ is the reconstructed \jpsi mass, $p_0$ is a 
normalization parameter, $p_4$ is the high mass tail parameter, and 
$p_1$, $p_2$ and $p_3$ are additional fit parameters.  It was found 
during the analysis that when fitting mass spectra with poor statistical 
precision, the shape of the correlated background was not well defined. 
This led to a contribution of less than 10\% to the point-to-point 
uncertainty in the \jpsi yields. Therefore, the shape of the correlated 
background as a function of \pt (determined by $p_1$, $p_2$ and $p_3$) 
was constrained using simulation results based on a detailed study of 
dimuon mass spectra~\cite{Aidala:2018ajl, Leung:2019vbb, 
Aidala:2019cnn,Adare:2010fn}. A systematic uncertainty on the \jpsi 
yield was assigned for this procedure by refitting the data with various 
combinations of correlated background parameters left free.

\subsection{Efficiency correction}

\subsubsection{Acceptance and Reconstruction Efficiency}
\label{sec:acc_eff}

The study of acceptance and reconstruction efficiency of dimuons from \jpsi 
decays has been performed using a \geant-based full detector 
simulation~\cite{G4}. In this simulation, the MuTr and MuID detector 
efficiencies are set to values determined from the data.  An emulator of the 
dimuon trigger response is included in the simulation to account for the 
trigger efficiency.  As these efficiencies depend on the instantaneous 
luminosity being sampled, each data set is divided into three groups with 
different beam interaction rates, and corrected yields with separate 
corrections are compared.  A systematic uncertainty is assigned to the 
extracted \jpsi cross sections times branching fraction to \mumu to reflect 
the differences, see Sec.~\ref{systunc} for details.

\begin{figure}[htb]
\includegraphics[width=0.96\linewidth]{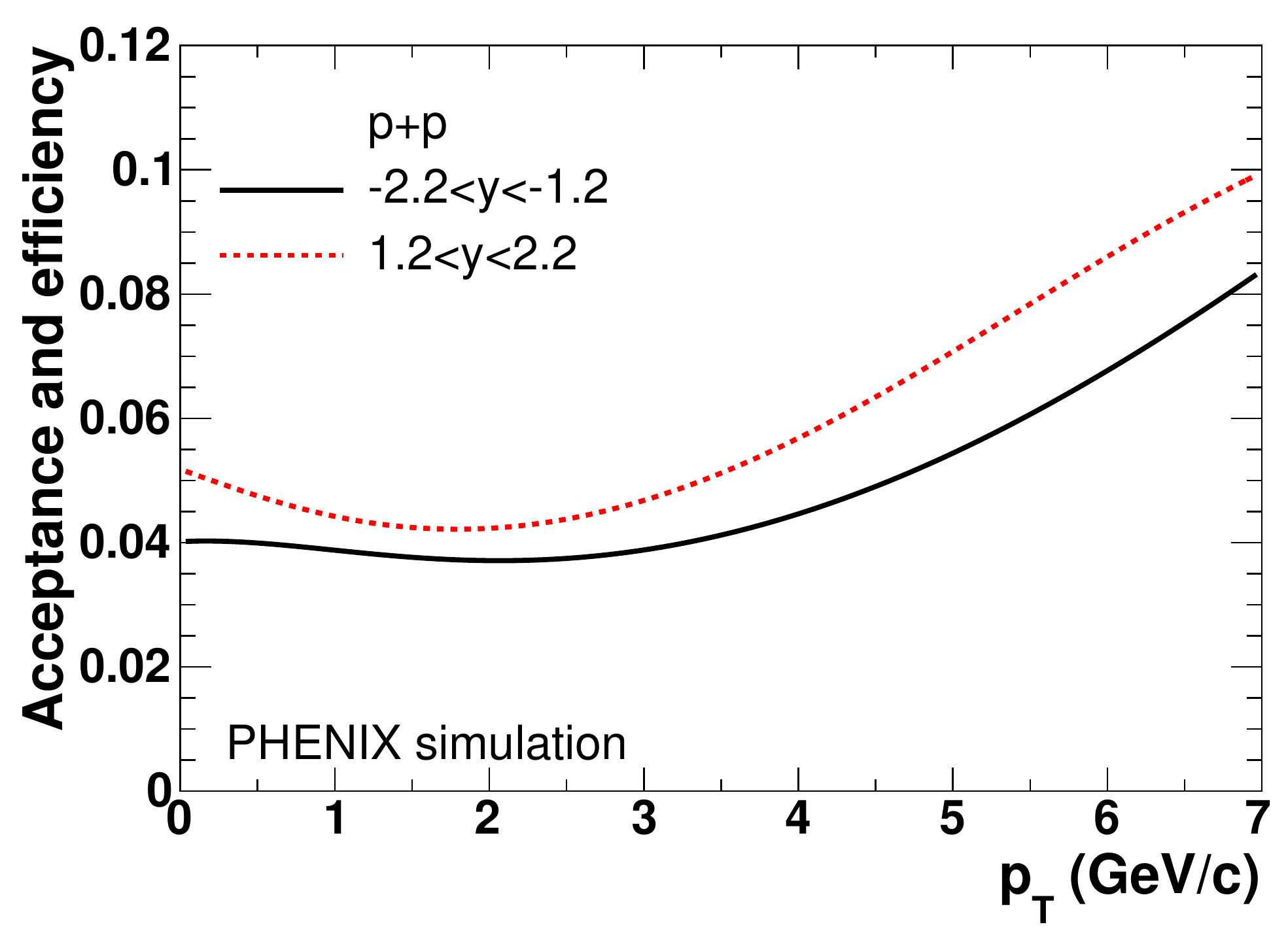}
\caption{\label{fig:acceff_pt}
Acceptance and reconstruction efficiency as a function of \pt for 
dimuons from \jpsi decays in \pp collisions.  GEANT simulations evaluate 
detector acceptance$\times$efficiency simultaneously.
}
\end{figure}

The \pythia event generator package~\cite{PYTHIA} is utilized to 
generate \jpsi events used for the full \textsc{Geant4} detector 
simulation. To take into account effects from background hits, the 
simulated hits of \pythia \jpsi events are embedded into real data 
events, separated into centrality classes of the collision system.  The 
track reconstruction is then run on the data with embedded simulated 
hits to examine the effects of the underlying event on the 
reconstruction efficiency. Figure~\ref{fig:acceff_pt} shows the 
acceptance and reconstruction efficiency for the \jpsi as a function of 
\pt in \pp collisions. The difference between the two muon arms is 
mainly from different inefficient detector areas. There is little 
multiplicity effect on the reconstruction efficiency in small collision 
systems, the relative difference between 0\%--20\% and 40\%--88\% 
centrality bins at backward rapidity in \heau collisions is about 5\%.

\subsection{Invariant yield and nuclear modification factor}

The invariant yield of dimuons from \jpsi decays in a given rapidity and 
centrality bin for the integrated \pt range is

\begin{equation}
\label{eq:yield_rap}
 B_{ll}\frac{dN}{dy} = \frac{1}{\Delta y} \frac{ c_\mathrm{BBC}}{ \varepsilon_\mathrm{Ae} \varepsilon_\mathrm{trig}}\frac{N_{J/\psi}}{N_\mathrm{evt}},    
\end{equation}
where $B_{ll}$ is the branching ratio of \jpsi to dimuons, $\Delta y$ is 
the width of the rapidity bin, $N_{J/\psi}$ is the number of \jpsi 
obtained from the fit procedure, $c_\mathrm{BBC}$ is the BBC bias 
correction factor described in Table~\ref{tab:centrality}, 
$N_\mathrm{evt}$ is the number of sampled MB events in the given 
centrality bin, $\varepsilon_\mathrm{Ae}$ is the \jpsi acceptance and 
reconstruction efficiency, and $\varepsilon_\mathrm{trig}$ is the dimuon 
trigger efficiency.

The invariant yield in a $y$, \pt, and centrality bin is

\begin{equation}
\label{eq:yield}
 \frac{B_{ll}}{2\pi p_T} \frac{d^{2}N}{dydp_T} = \frac{1}{2\pi \pt 
\Delta\pt \Delta y} \frac{c_\mathrm{BBC}}{\varepsilon_\mathrm{Ae} 
\varepsilon_\mathrm{trig}} \frac{N_{J/\psi}}{N_\mathrm{evt}},
\end{equation}
where $\Delta \pt$ is the width of the \pt bin, and in this case 
$N_\mathrm{evt}$ is the number of events in the centrality bin.
Based on the invariant yields calculated with Eq.~\ref{eq:yield}, the 
\jpsi nuclear modification factor $R_{AB}$ for a given $y$, \pt, and 
centrality bin is formed to quantify nuclear effects in \pal, \pau, and 
\heau collisions. The $R_{AB}$ is defined as

\begin{equation}
\label{eq:rab}
    R_{AB} = \frac{1}{\langle \ncoll 
\rangle}\frac{d^{2}N^{AB}/dydp_T}{d^{2}N^{pp}/dydp_T},
\end{equation}
where $d^2N^{AB}/dyd\pt$ is the \jpsi invariant yield for a certain 
centrality bin of $A$$+$$B$ collisions, $d^2N^{pp}/dyd\pt$ is the 
corresponding \jpsi invariant yield for \pp collisions, and $\langle 
\ncoll \rangle$ is the mean number of binary collisions for that 
centrality bin in $A$$+$$B$ collisions.

\subsection{\meanptsq calculation}

The \meanptsq values for various centrality bins in all collision 
systems have been calculated over the full measured \pt range ($0<\pt<7$ 
GeV/$c$). We do not extrapolate the \pt distribution beyond 7 GeV/$c$. A 
previous study~\cite{Adare:2012qf} determined that extrapolating to 
infinite \pt increased the \meanptsq values by 3\%. The value of 
\meanptsq is calculated numerically using the following formula:

\begin{equation}
    \meanptsq = \frac{\sum \limits^{N}_{i=0}p_{T,i}^2 w_{i}}{\sum 
\limits^{N}_{i=0} w_{i}},
\end{equation}
where $p_{T,i}$ is the center of the i-\textit{th} \pt bin, and $w_i$ is 
the weight factor proportional to the \jpsi invariant yield in the \pt 
bin:

\begin{equation}
\label{eq:weight}
    w_i = p_{T,i} dp_{T,i} \left( \frac{B_{ll}}{2\pi p_T} 
\frac{d^{2}N}{dydp_T} \right)_i,
\end{equation}
where $dp_{T,i}$ is the width of the bin.

\subsection{Systematic Uncertainties}
\label{systunc}

In the measurements we present in the next section, Type A uncertainties 
are uncorrelated point to point uncertainties, and are dominated by the 
statistical precision of the data.  Type B systematic uncertainties are 
correlated point to point uncertainties.  Type C global uncertainties 
are fractional uncertainties that apply to all measurements uniformly.

\subsubsection{Signal extraction}

As discussed in Sec.~\ref{sec:background}, the modified Hagedorn 
function in Eq.~\ref{eq:hagedorn} was used to describe the correlated 
background. Initial parameters were estimated based on the previous 
measurement of dimuon mass spectra~\cite{Aidala:2018ajl, Leung:2019vbb}, 
and two parameters, $p_0$ and $p_4$, were left free to describe dimuon 
mass distributions in the data more properly. For the systematic 
uncertainty study, additional parameters, $p_1$, $p_2$, and $p_3$, in 
the modified Hagedorn function were also freed in the fit procedure. We 
observe 1.4\%--2.8\% variations of \jpsi counts depending on rapidity, 
\pt, and centrality.

To describe the combinatorial background shape, the modified Hagedorn 
function in Eq.~\ref{eq:hagedorn}, used for the correlated background 
component, was also used to fit like-sign dimuon mass distributions. The 
effect of statistical fluctuations in the like-sign dimuon mass 
distributions was studied by varying the shape based on the statistical 
uncertainties of the fit parameters. We observe 1.0\%--4.4\% variations 
of \jpsi counts depending on rapidity, \pt, and centrality.

The uncertainty related to fixing the \jpsi mass centroid and width was 
evaluated by directly comparing the difference in yields with the 
parameters free versus fixed, which ranges from 1.1\%--2.9\% 
uncertainty.

Table II lists all Type B uncertainties arising from the \jpsi signal 
extraction.

\begin{table}[tbh]
\caption{\label{tab:sys_fit}
Fractional systematic uncertainties on the signal extraction in \pp, 
\pal, \pau, and \heau collisions at forward (north arm) and backward 
(south arm) rapidity.
}
\begin{ruledtabular}
\begin{tabular}{ccccc}
 System & Source        & Forward       & Backward     & Type\\
\hline
\pp   & Corr. bkg.      & 1.4\%         & 1.8\%        & B   \\ 
\pal  &                 & 1.4\%         & 1.8\%        & B   \\
\pau  &                 & 1.9\%--2.7\%  & 1.4\%--2.8\% & B   \\
\heau &                 & 2.3\%--2.4\%  & 1.4\%--2.8\% & B   \\
\\
\pp   & Comb. bkg.      & $<$1.0\%      & $<$1.0\%     & B   \\ 
\pal  &                 & 1.0\%         & 4.4\%        & B   \\
\pau  &                 & 1.0\%         & 1.0\%        & B   \\
\heau &                 & 1.0\%         & 2.7\%        & B   \\ 
\\
\pp   & Signal shape    & -             &  -           & B   \\ 
\pal  &                 & 1.1\%         & 1.1\%        & B   \\
\pau  &                 & 0\%--1.5\%    & 0\%--2.9\%   & B   \\
\heau &                 & 1.5\%         & 2.9\%        & B 
\end{tabular}
\end{ruledtabular}
\end{table}

\subsubsection{Acceptance and efficiency correction}

The acceptance and reconstruction efficiency correction and trigger 
efficiency correction are obtained from simulation, so discrepancies 
between the data and calculations can be a source of systematic uncertainty. The 
discrepancies can be due to a variation in the detector performance 
during the data taking period and/or inaccuracy of detector geometry and 
dead channel maps in the simulation. To quantify these effects, 
we divide each data set into three groups of different detector 
efficiency, based on the beam instantaneous luminosity and calculated 
invariant yields with separate correction factors. In this comparison we 
observe 1.5\%--5.0\% variations, depending on rapidity and data set, and 
assign this variation as a systematic uncertainty. In addition, we 
compare the azimuthal angle $\phi$ distribution of tracks in the MuTr 
between the data and simulation, and assign a 2.5\%--6.0\% systematic 
uncertainty depending on rapidity and data set.

In the simulation procedure, \pythia was used to generate \jpsi events, 
and initial \jpsi rapidity and \pt shapes in \pythia are tuned to match 
the measurements in \pp and \dau 
collisions~\cite{Adare:2011vq,Adare:2012qf,Adare:2010fn}.  These two 
different assumptions of the distributions are used as bounds to 
estimate the sensitivity of this analysis to the shapes of these 
distributions in \pal, \pau, and \heau collisions, which are not known 
$a$ $priori$. The variation of acceptance and reconstruction efficiency 
between two sets of rapidity and \pt distributions is less than 2\%, so 
we assigned a 2\% conservative systematic uncertainty.

The uncertainty in the dimuon acceptance caused by lack of knowledge of 
the \jpsi polarization was studied as described in~\cite{Adare:2011vq}. 
Because there is no precise measurement of \jpsi polarization, a maximum 
polarization value ($\pm1$ in the helicity frame) was considered to 
study the systematic uncertainty. The variation of dimuon acceptance 
becomes larger as \jpsi \pt decreases, and 9\%--20\% systematic 
uncertainties are assigned depending on \pt. We assumed that the \jpsi 
polarization is not significantly modified in \pal, \pau, and \heau 
collisions, and this uncertainty is canceled in the \rab calculation.  
This assumption was also made in a similar PHENIX analysis for \jpsi 
nuclear modification in $d$$+$Au collisions~\cite{Adare:2012qf}.

To evaluate a systematic uncertainty on the dimuon trigger 
efficiency, the single muon trigger efficiency in the MB triggered data 
obtained with a large number of muon samples was compared with the 
emulated single muon trigger efficiency determined from simulation. This 
difference was propagated to the uncertainty in the dimuon trigger 
efficiency based on a previous study~\cite{Aidala:2018ajl}, and a 
1.0\%--4.8\% systematic uncertainty was assigned.  The Type B systematic 
uncertainties related to acceptance and efficiency correction are shown 
in Table III.

\begin{table}[tbh]
\caption{\label{tab:sys_acceff}
Fractional systematic uncertainties on the acceptance and efficiency 
correction in \pp, \pal, \pau and \heau collisions at forward (north 
arm) and backward (south arm) rapidity.
}
\begin{ruledtabular} \begin{tabular}{ccccc}
System & Source          & Forward        & Backward        & Type\\\hline
\pp    & Run variation     & 4.0\%   & 4.7\%    & B   \\
\pal   &     & 2.8\%   & 3.3\%    & B   \\
\pau   &     & 1.6\%   & 3.5\%    & B   \\
\heau  &     & 1.5\%   & 5.0\%    & B   \\
\\
\pp    &$\phi$ Matching & 5.8\%   & 5.0\%    & B   \\ 
\pal   & & 3.6\%   & 3.3\%    & B   \\
\pau   & & 3.4\%   & 4.0\%    & B   \\
\heau  & & 3.1\%   & 2.5\%    & B   \\
\\
all    & Initial shape  & 2.0\%   & 2.0\%    & B   \\
\\
all    & \jpsi pol.     & 10\%--20\%     & 9\%--20\%       & B   \\ 
\\
\pp    & Trigger eff.     & 1.0\%--1.7\%     & 1.0\%--2.6\%      & B   \\ 
\pal   &      & 1.0\%--1.8\%     & 2.0\%--4.6\%      & B   \\
\pau   &      & 1.0\%--1.7\%     & 1.0\%--4.8\%      & B   \\
\heau  &      & 1.0\%--2.4\%     & 1.0\%--2.4\%      & B   \\
\end{tabular} \end{ruledtabular}
\end{table}

\begin{figure}[tbh]
\includegraphics[width=0.96\linewidth]{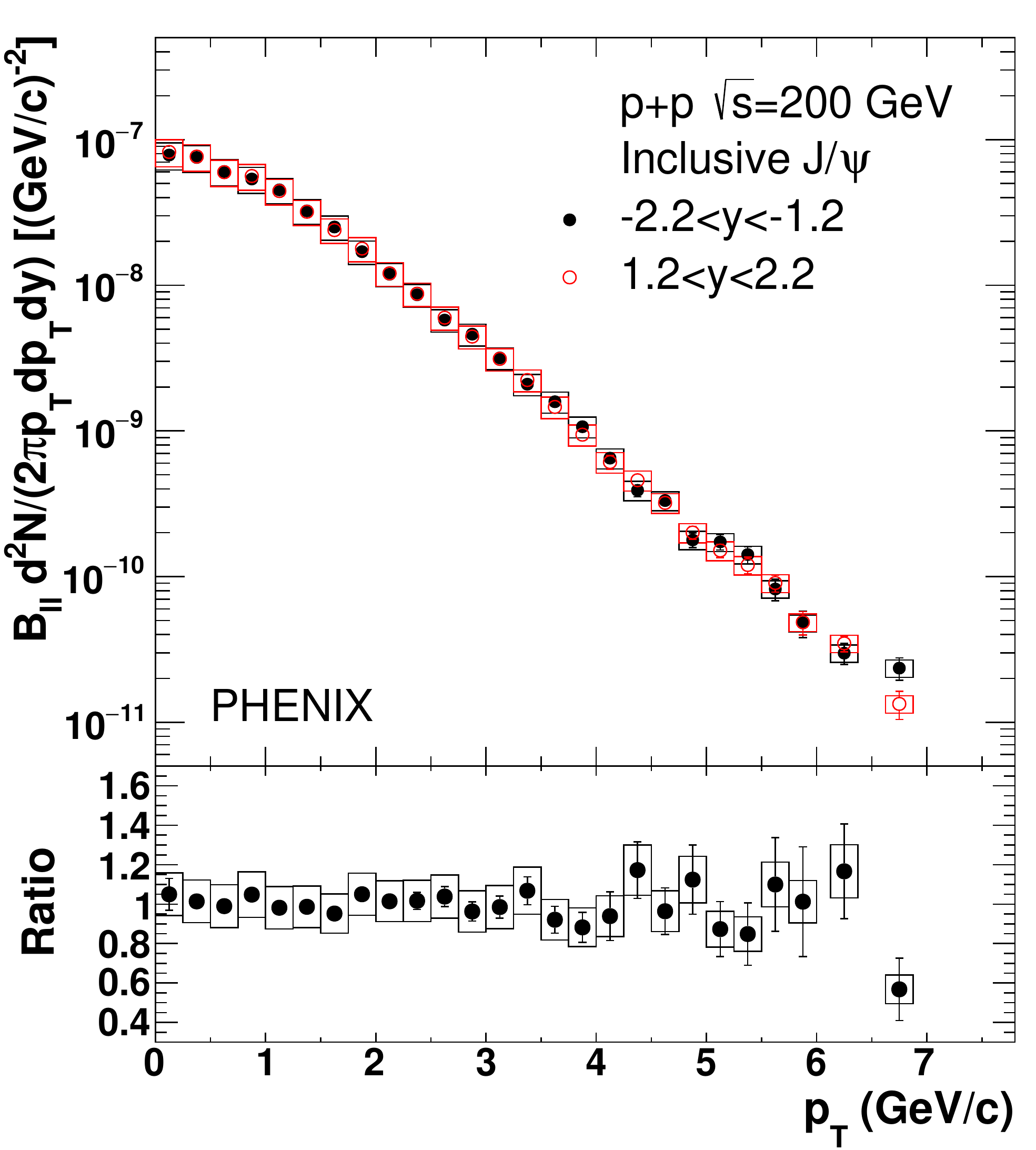}
\caption{\label{fig:dndpt_pp}
\jpsi invariant yields as a function of \pt in \pp collisions at 
$\sqrt{s}=200~\mathrm{GeV}$. The ratio between the values for the two 
muon arms is presented in the bottom panel. Bars (boxes) around data 
points represent point-to-point uncorrelated (correlated) uncertainties. 
There is also a global systematic uncertainty of 10.1\%}
\end{figure}

\subsubsection{Multiple interaction}

Due to the high instantaneous beam luminosity, particularly in \pp and 
\pal runs, it is possible to have multiple inelastic collisions from a 
single beam crossing, which can affect the invariant yield calculation. 
To investigate this effect, the variation among invariant 
yields in three groups of different instantaneous luminosity for each 
data set was studied, revealing a yield variation smaller than 5\%. 
However, the instantaneous luminosity dependence of the acceptance and 
efficiency correction is already included as a systematic uncertainty, 
and so no additional systematic uncertainty is assigned.

\subsubsection{\meanptsq}

The \meanptsq uncertainty is calculated based on the systematic 
uncertainty of the invariant yield as a function of \pt. The systematic 
uncertainties are mostly point-to-point correlated, and we assumed that 
the uncertainties in different \pt bins are linearly correlated. The 
upper and lower limits of invariant yield in each \pt bin are taken to 
calculate the upper and lower limits of \meanptsq.

\subsubsection{\meanncoll and BBC efficiency}

The systematic uncertainties on the BBC efficiency and the determination 
of \meanncoll in \pal, \pau, and \heau collisions described in 
Table~\ref{tab:centrality} are evaluated by following the procedure 
developed in the previous PHENIX analyses of \dau 
data~\cite{Adare:2013nff}.
These systematic uncertainties are considered as Type C (Type B) systematic uncertainties in rapidity and \pt (centrality) dependence results. 
The systematic uncertainty on the BBC 
efficiency in \pp collisions obtained in~\cite{Adler:2003pb} is 
$10.1\%$, and this systematic uncertainty is considered as a Type C systematic uncertainty.

\section{Results}

\begin{figure*}[tbh] 
\begin{minipage}{0.98\linewidth}
\includegraphics[width=1.0\linewidth]{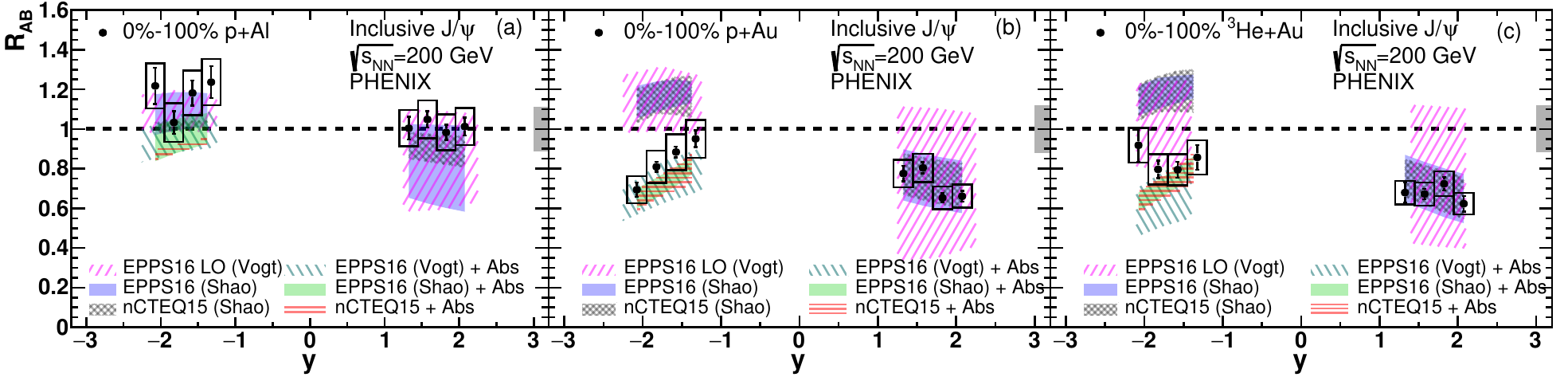} 
\caption{\label{fig:rab_y_mb} Nuclear modification factor of inclusive 
\jpsi as a function of rapidity for 0\%--100\% \pal (a), \pau (b), and 
\heau (c) collisions. Bars (boxes) around data points represent 
point-to-point uncorrelated (correlated) uncertainties. The theory bands 
are discussed in the text.}
\end{minipage}
\begin{minipage}{0.98\linewidth}
\includegraphics[width=1.0\linewidth]{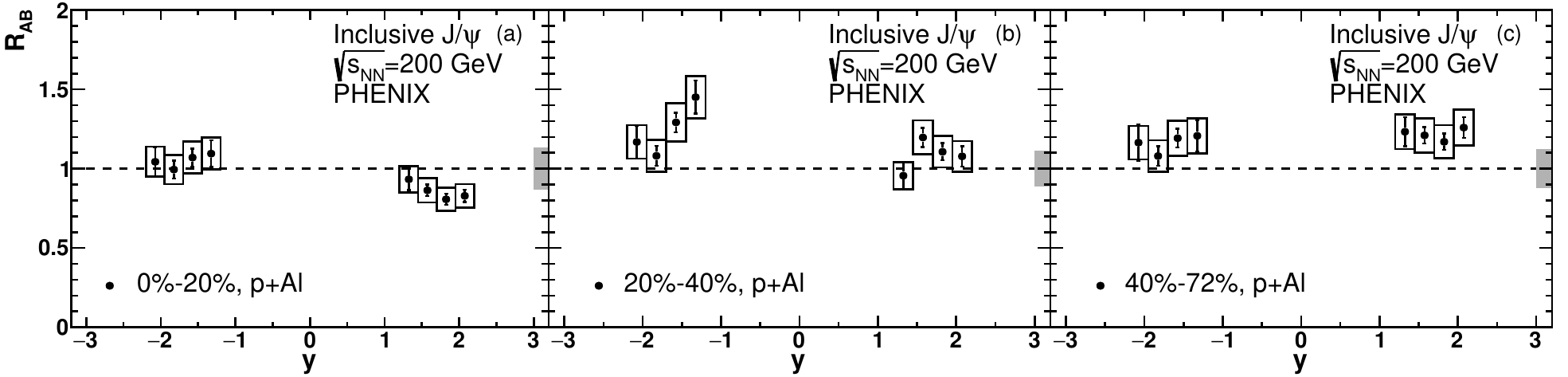}
\caption{\label{fig:rab_y_pal_cent}
Nuclear modification factor of inclusive \jpsi as a function of rapidity 
in three centrality bins for \pal collisions. Bars (boxes) around data 
points represent point-to-point uncorrelated (correlated) 
uncertainties.}
\end{minipage}
\begin{minipage}{0.98\linewidth}
\includegraphics[width=1.0\linewidth]{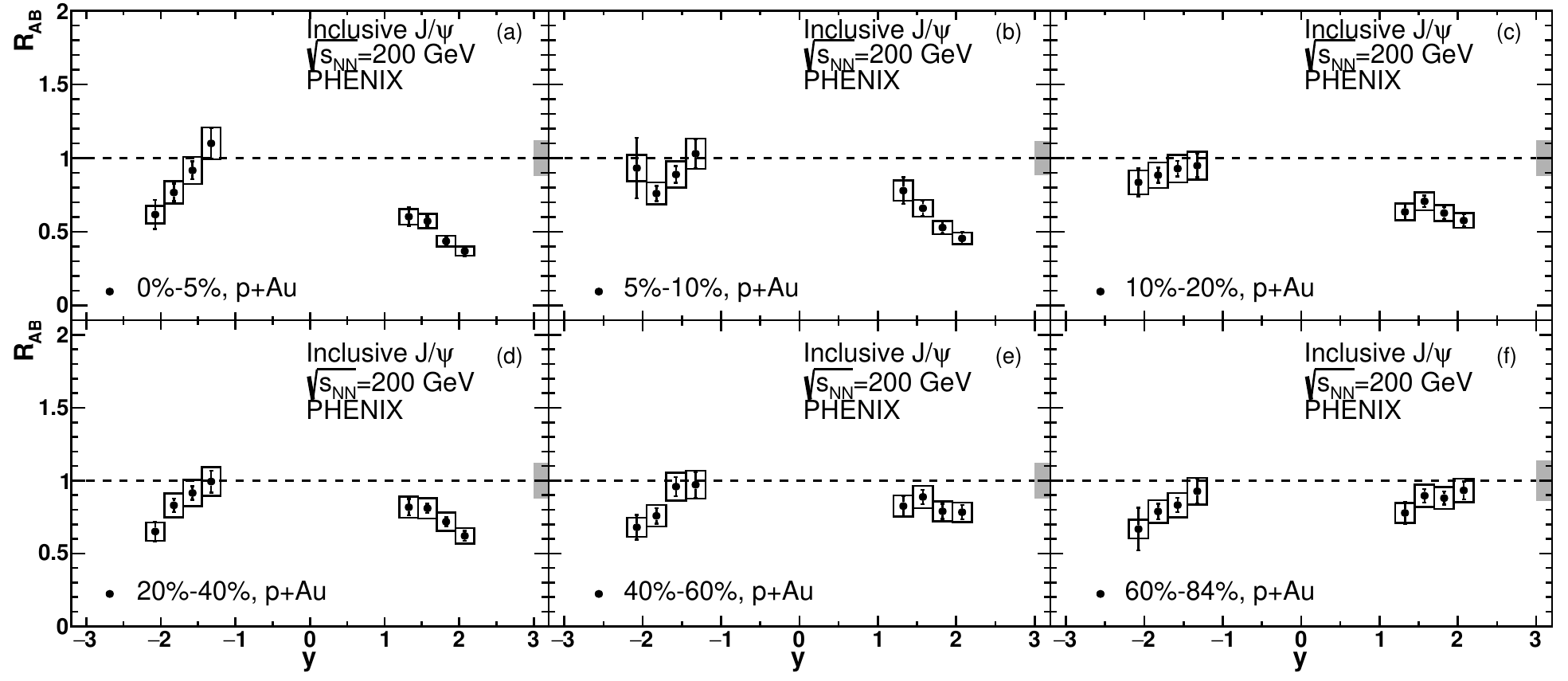}
\caption{\label{fig:rab_y_pau_cent}
Nuclear modification factor of inclusive \jpsi as a function of rapidity 
in six centrality bins for \pau collisions. Bars (boxes) around data 
points represent point-to-point uncorrelated (correlated) 
uncertainties.}
\end{minipage}
\begin{minipage}{0.98\linewidth}
\includegraphics[width=1.0\linewidth]{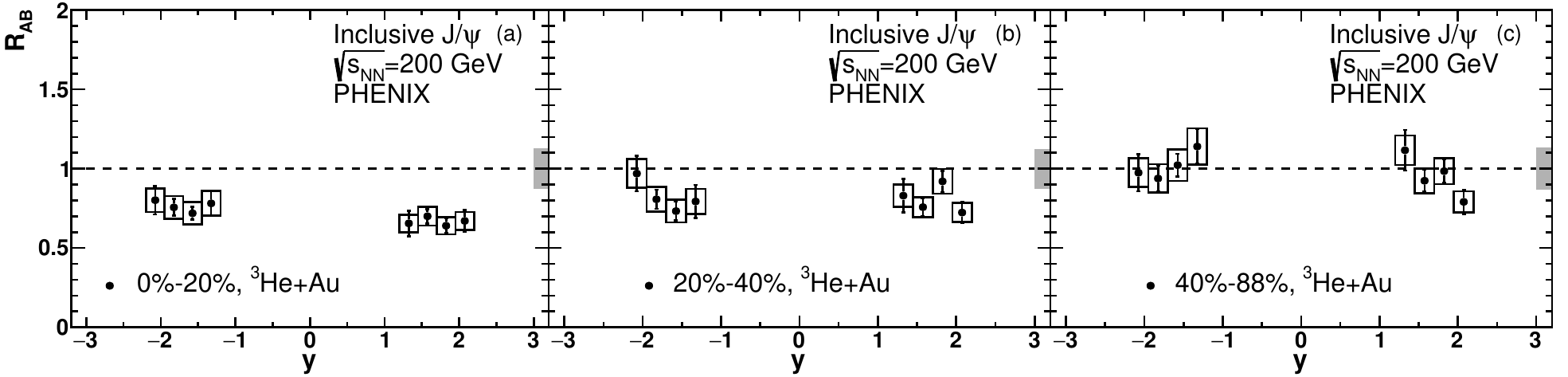}
\caption{\label{fig:rab_y_heau_cent}
Nuclear modification factor of inclusive \jpsi as a function of rapidity 
in three centrality bins for \heau collisions. Bars (boxes) around data 
points represent point-to-point uncorrelated (correlated) 
uncertainties.}
\end{minipage}
\end{figure*}

\begin{figure*}[htb]
\begin{minipage}{0.98\linewidth}
\includegraphics[width=1.0\linewidth]{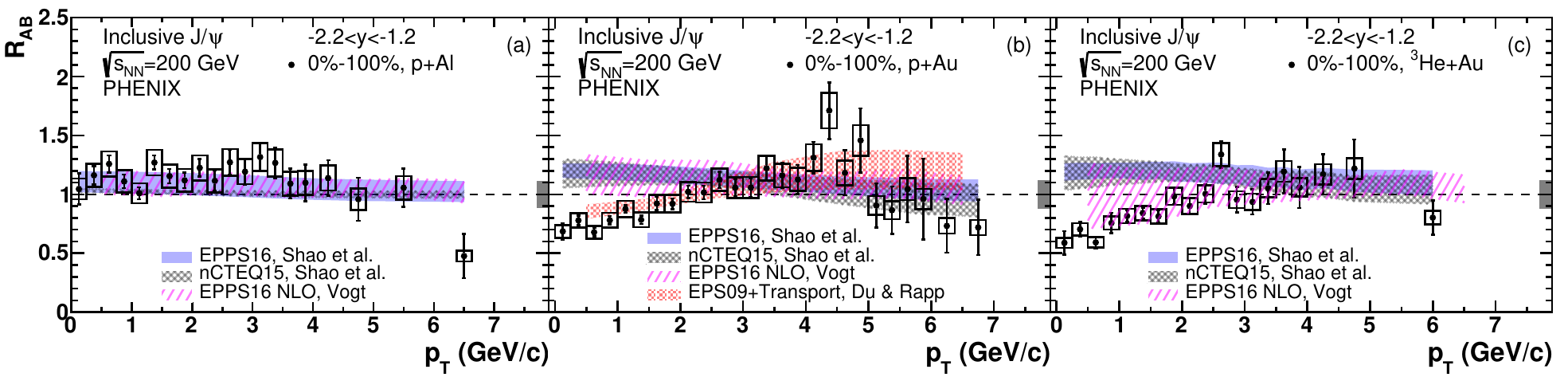}
\caption{\label{fig:rab_pt_mb_bwd}
Nuclear modification factor of inclusive \jpsi as a function of \pt at 
backward rapidity (Al/Au-going direction) for 0\%--100\% \pal, \pau, and 
\heau collisions. Bars (boxes) around data points represent 
point-to-point uncorrelated (correlated) uncertainties. The theory bands 
are discussed in the text.}
\end{minipage}
\begin{minipage}{0.98\linewidth}
\includegraphics[width=1.0\linewidth]{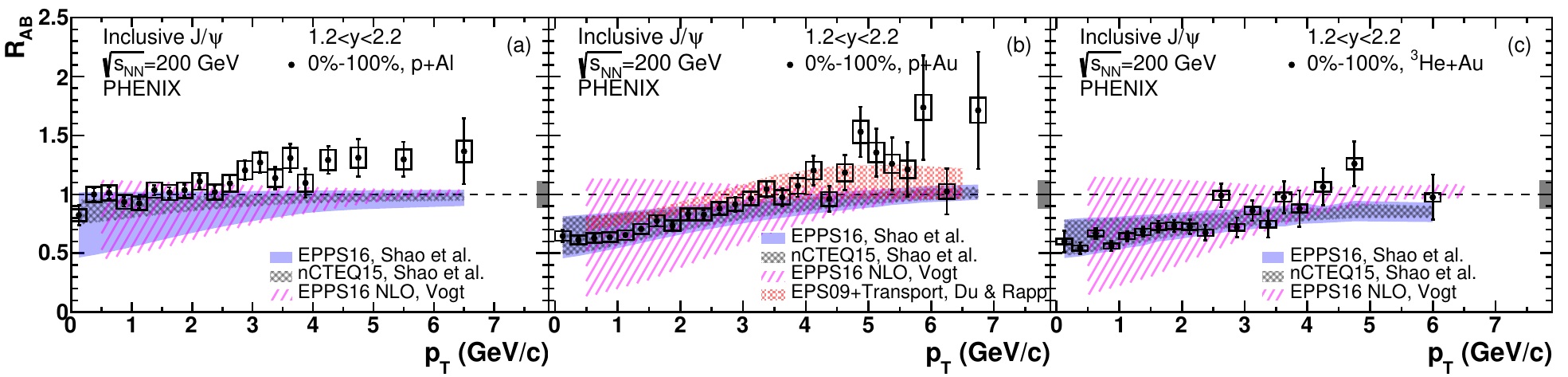}
\caption{\label{fig:rab_pt_mb_fwd}
Nuclear modification factor of inclusive \jpsi as a function of \pt at 
forward rapidity ($p$/$^{3}$He-going direction) for 0\%--100\% \pal, 
\pau, and \heau collisions. Bars (boxes) around data points represent 
point-to-point uncorrelated (correlated) uncertainties. The theory bands 
are discussed in the text.}
\end{minipage}
\begin{minipage}{0.98\linewidth}
\includegraphics[width=1.0\linewidth]{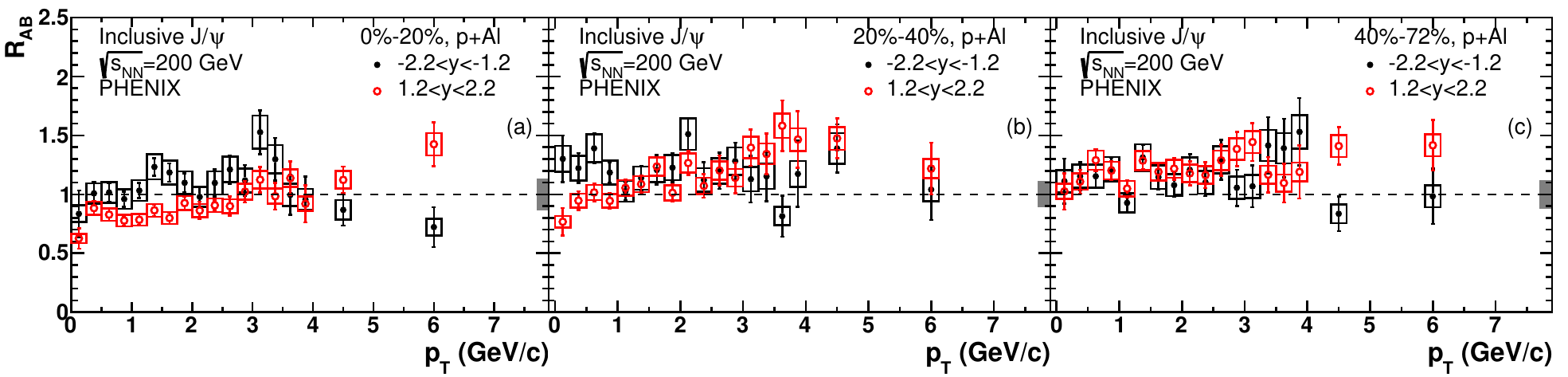}
\caption{\label{fig:rab_pt_pal_cent}
Nuclear modification factor of inclusive \jpsi as a function of \pt in 
three centrality bins for \pal collisions. Bars (boxes) around data 
points represent point-to-point uncorrelated (correlated) 
uncertainties.}
\end{minipage}
\end{figure*}

\begin{figure*}[htb]
\begin{minipage}{0.98\linewidth}
\includegraphics[width=1.0\linewidth]{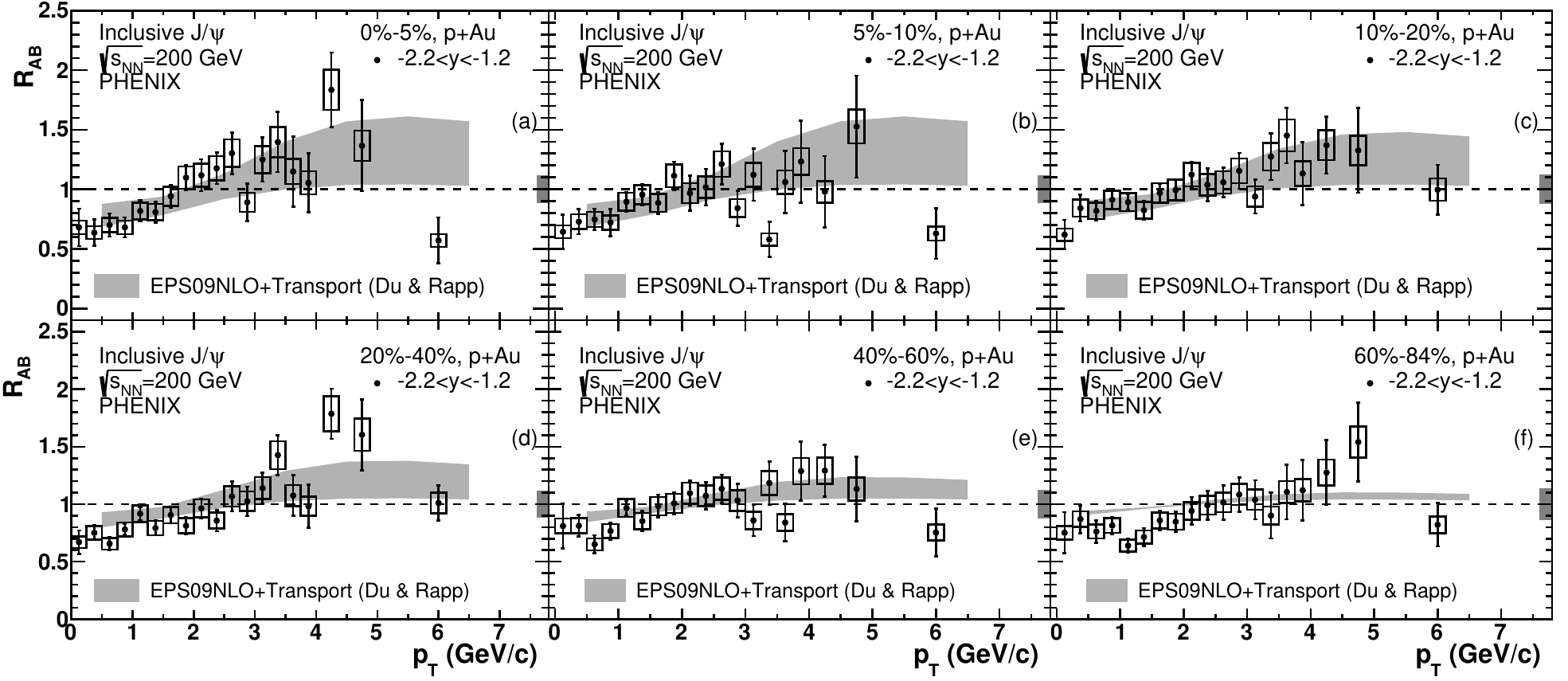}
\caption{\label{fig:rab_pt_pau_cent_arm0}
Nuclear modification factor of inclusive \jpsi as a function of \pt at 
$-2.2<y<-1.2$ in six centrality bins for \pau collisions. Bars (boxes) 
around data points represents point-to-point uncorrelated (correlated) 
uncertainties.  The theory bands are discussed in the text.  Note that 
the theory bands compared with the 0\%--5\% and 5\%--10\% centrality 
data are for 0\%--10\%.}

\vspace{5pt}
\end{minipage}
\begin{minipage}{0.98\linewidth}
\includegraphics[width=1.0\linewidth]{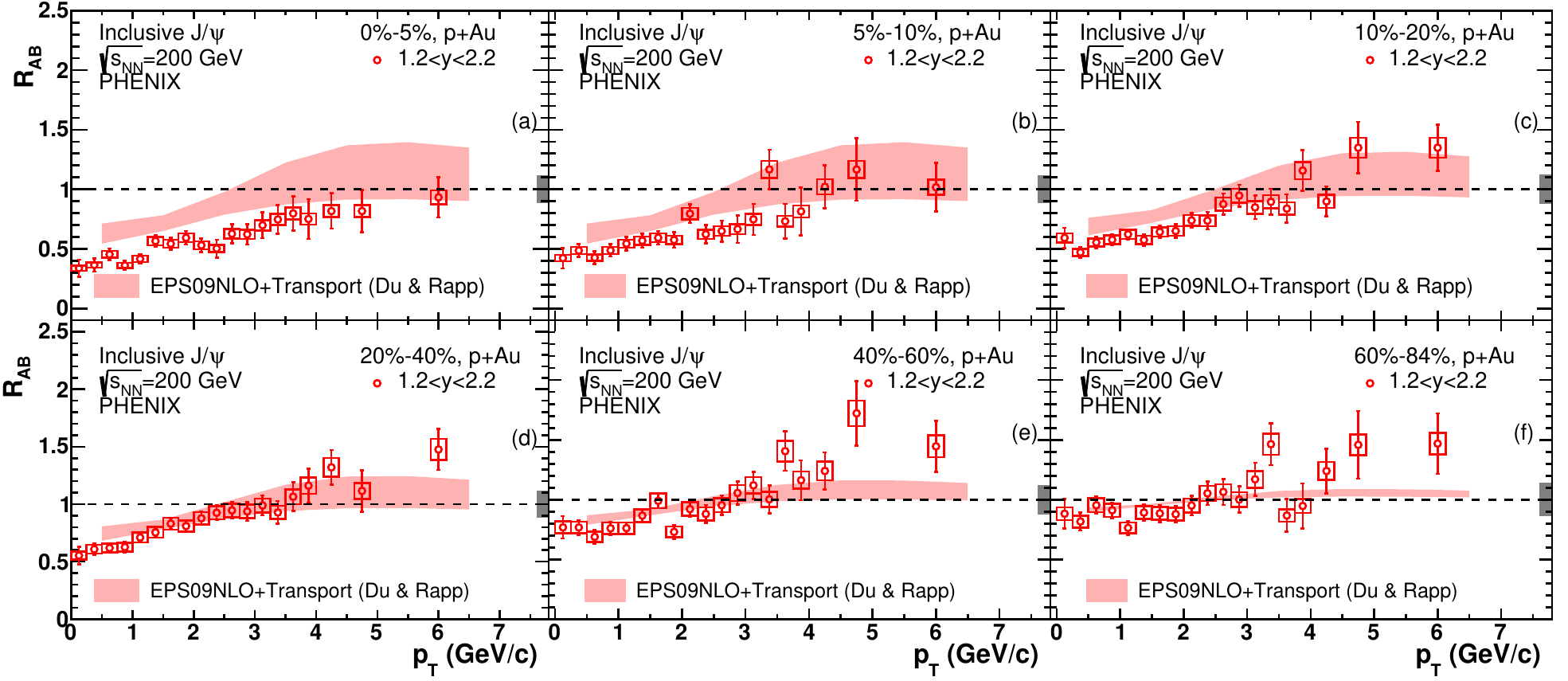}
\caption{\label{fig:rab_pt_pau_cent_arm1}
Nuclear modification factor of inclusive \jpsi as a function of \pt at 
$1.2<y<2.2$ in six centrality bins for \pau collisions. 
The theory bands are discussed in the text. Note that 
the theory bands compared with the 0\%--5\% and 5\%--10\% centrality 
data are for 0\%--10\%.}

\vspace{5pt}
\end{minipage}
\begin{minipage}{0.98\linewidth}
\includegraphics[width=1.0\linewidth]{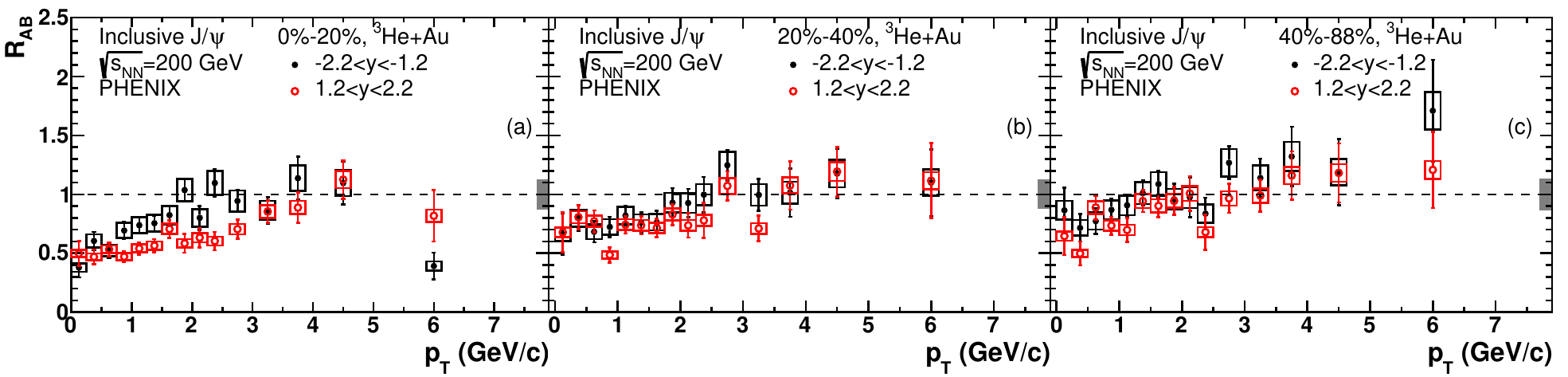}
\caption{\label{fig:rab_pt_heau_cent}
Nuclear modification factor of inclusive \jpsi as a function of \pt in 
three centrality bins for \heau collisions. 
}
\end{minipage}
\end{figure*}

\begin{figure*}[htb]
\includegraphics[width=1.0\linewidth]{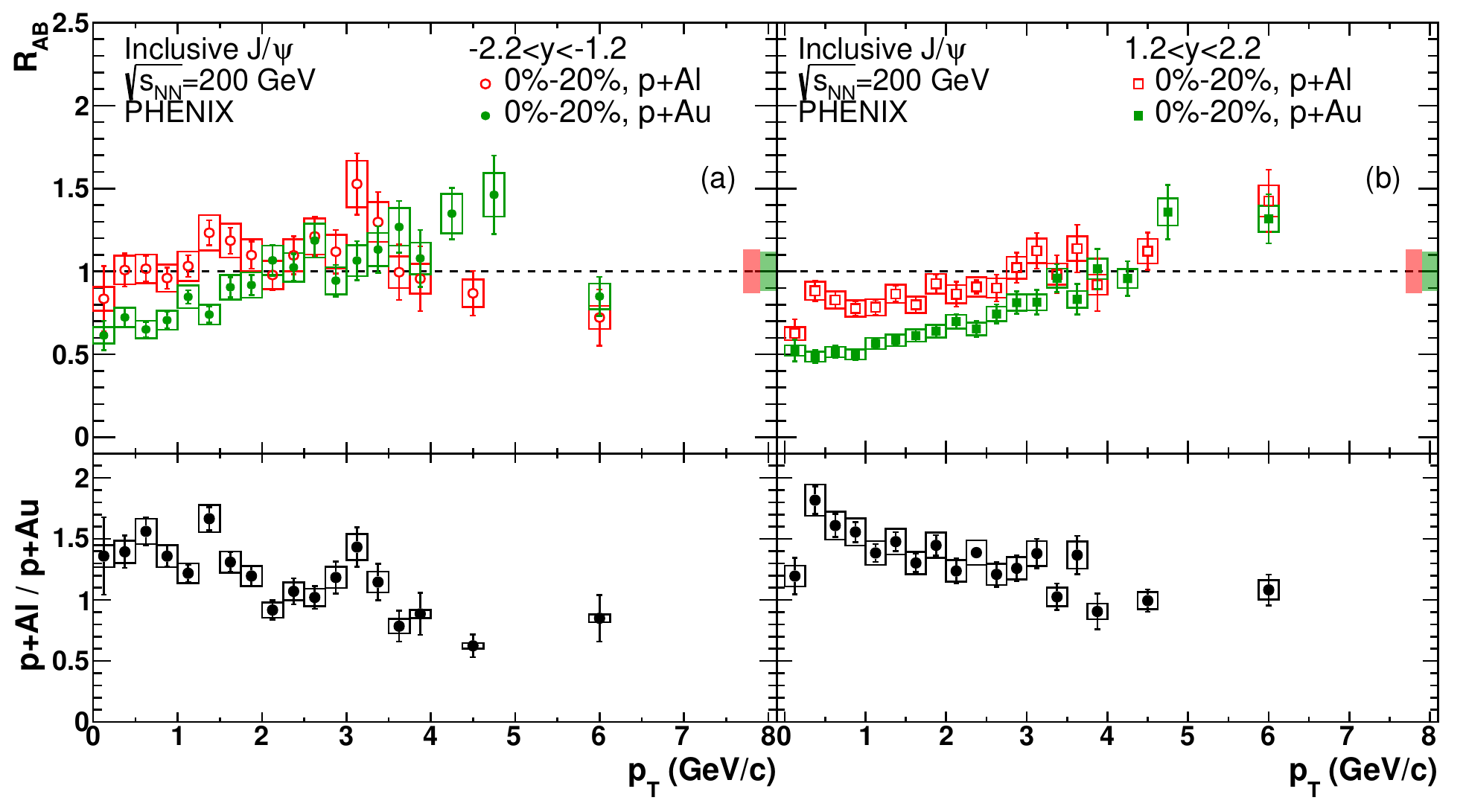}
\caption{\label{fig:rab_pt_pal_pau_0020_fwd}
Comparison of nuclear modification factor of \jpsi as a function of \pt 
in 0\%--20\% centrality \pal and \pau collisions. Bars (boxes) around 
data points represent point-to-point uncorrelated (correlated) 
uncertainties.}
\end{figure*}

\begin{figure*}[htb]
\includegraphics[width=1.0\linewidth]{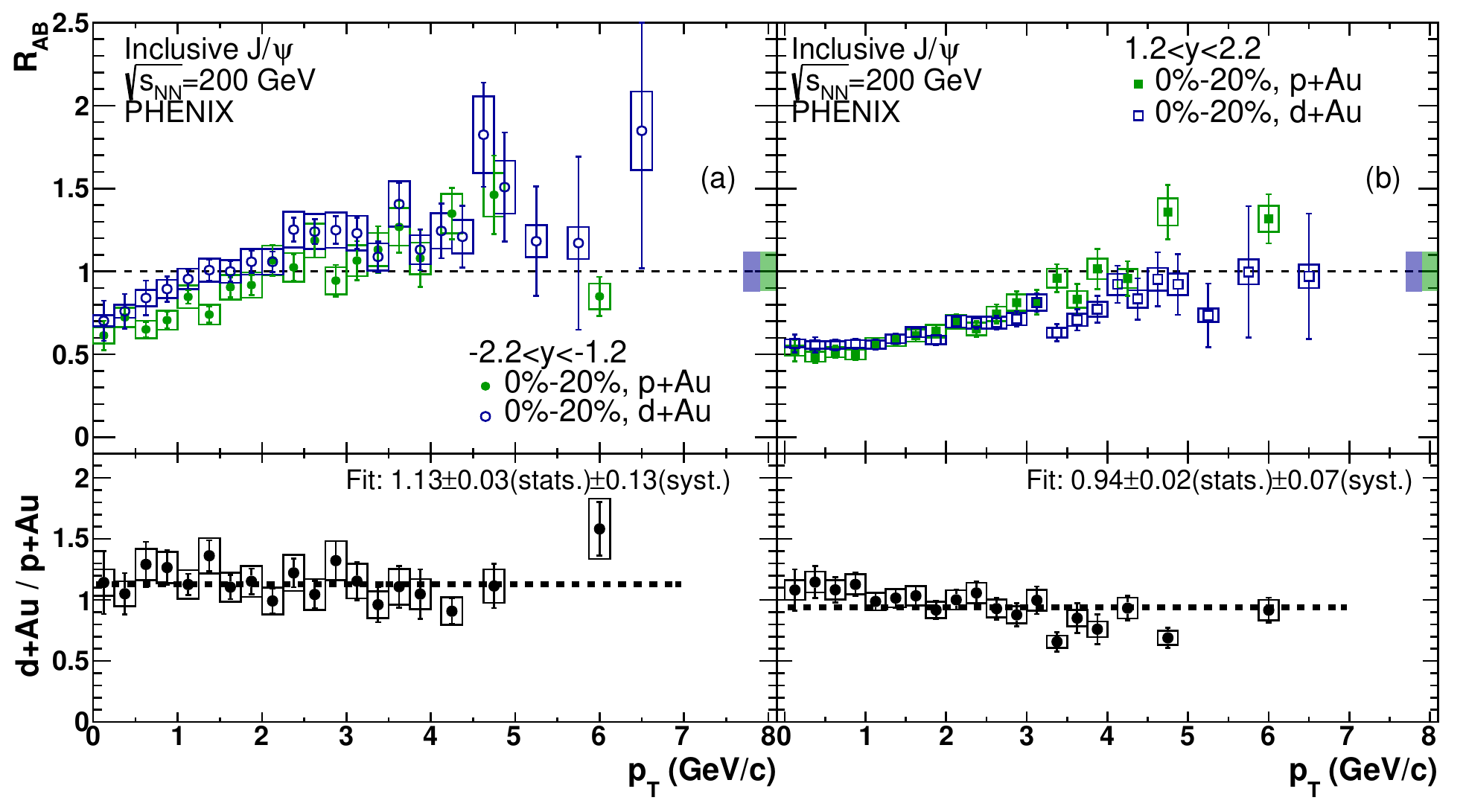}
\caption{\label{fig:rab_pt_dAu_pau_0020_fwd}
Comparison of nuclear modification factor of \jpsi as a function of \pt 
in 0\%--20\% centrality $d$$+$Au~\protect\cite{Adare:2012qf} and 
\pau collisions. Bars (boxes) around data points represent point-to-point 
uncorrelated (correlated) uncertainties.}
\end{figure*}

\begin{figure*}[htb]
\includegraphics[width=1.0\linewidth]{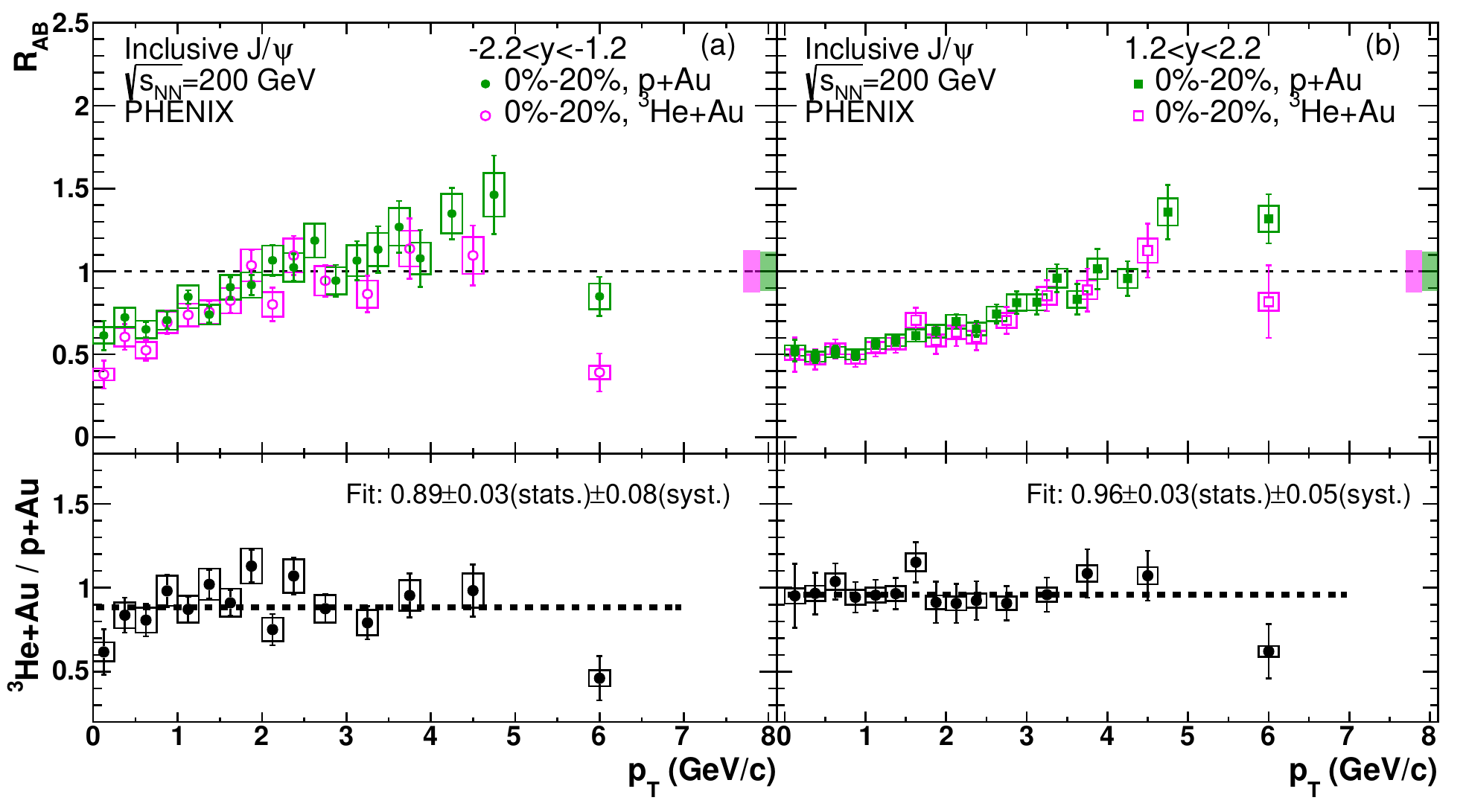}
\caption{\label{fig:rab_pt_pau_heau_0020_fwd}
Comparison of nuclear modification factor of \jpsi as a function of \pt 
in 0\%--20\% centrality \pau and \heau collisions. Bars (boxes) around 
data points represent point-to-point uncorrelated (correlated) 
uncertainties.}
\end{figure*}

\begin{figure*}[htb]
\includegraphics[width=1.0\linewidth]{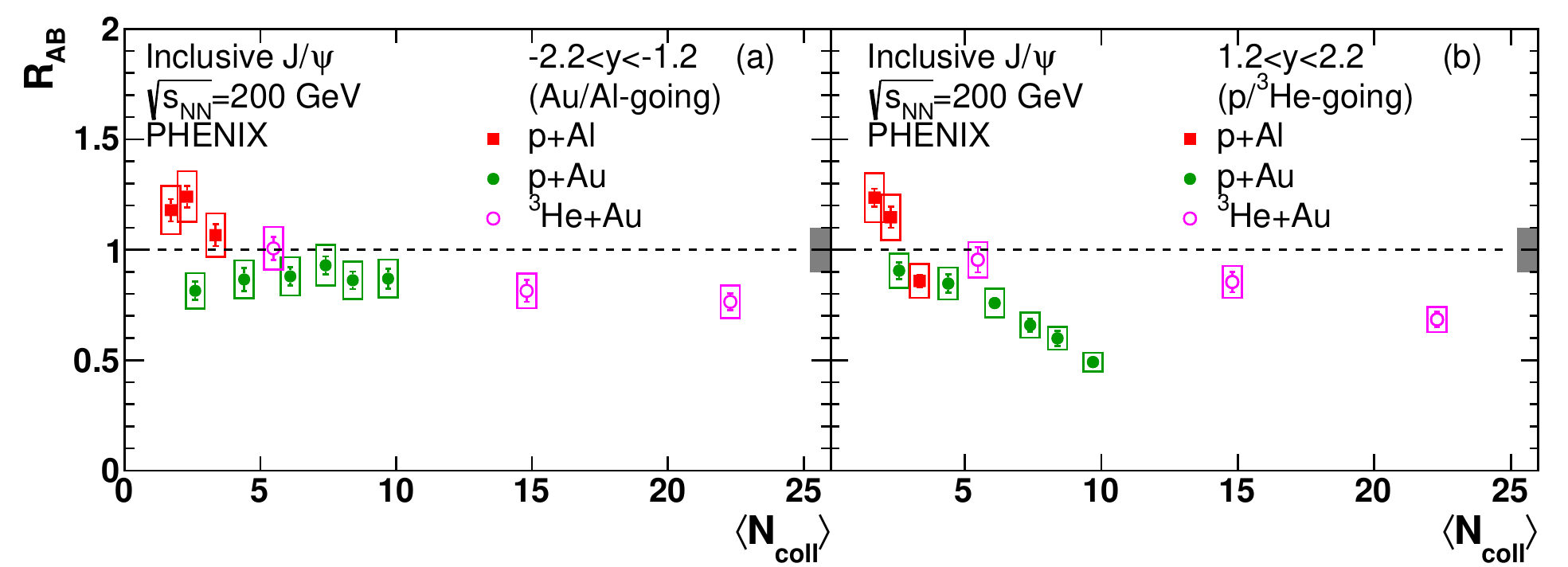}
\caption{\label{fig:rab_ncoll}
Nuclear modification factor of \jpsi as a function of $\langle \Ncoll 
\rangle$ for \pal, \pau, and \heau collisions. Bars (boxes) around data 
points represent point-to-point uncorrelated (correlated) 
uncertainties.}
\end{figure*}

\begin{figure*}[htb]
\includegraphics[width=1.0\linewidth]{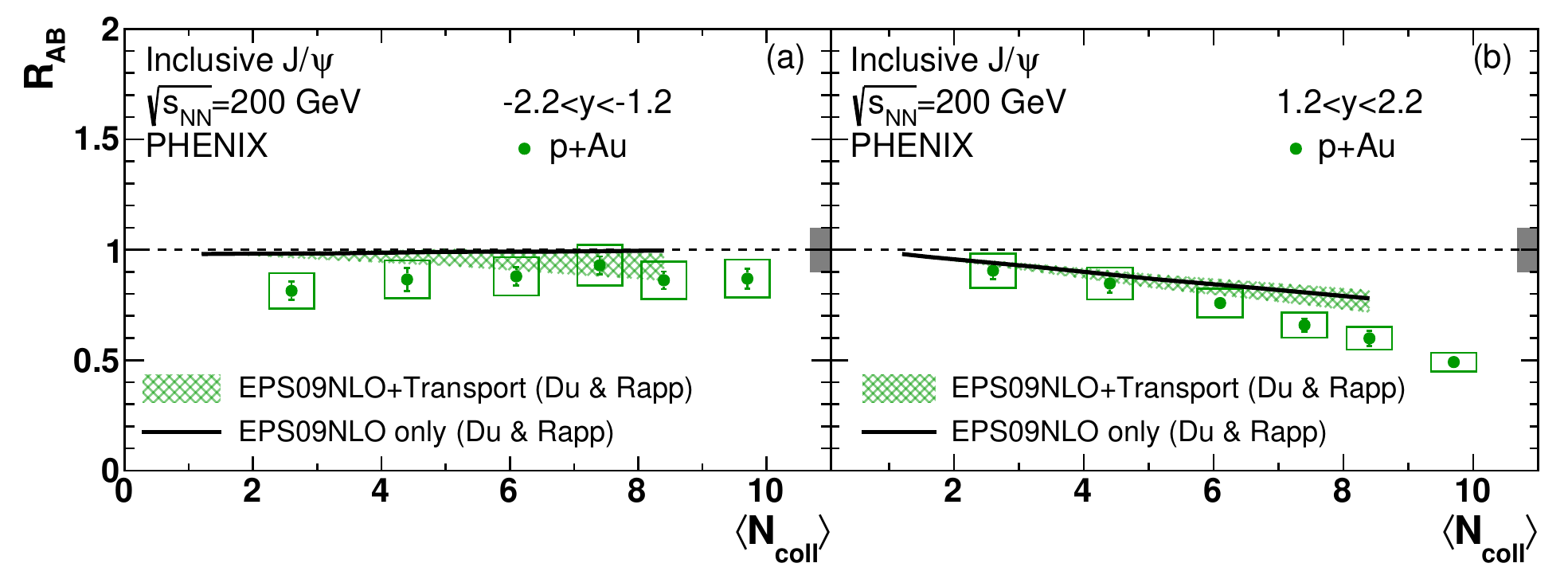}
\caption{\label{fig:rab_ncoll_pAu}
Nuclear modification factor of \jpsi as a function of $\langle \Ncoll 
\rangle$ for \pau collisions compared with the transport model. Bars 
(boxes) around data points represent point-to-point uncorrelated 
(correlated) uncertainties.}
\end{figure*}

\begin{figure*}[htb]
\includegraphics[width=1.0\linewidth]{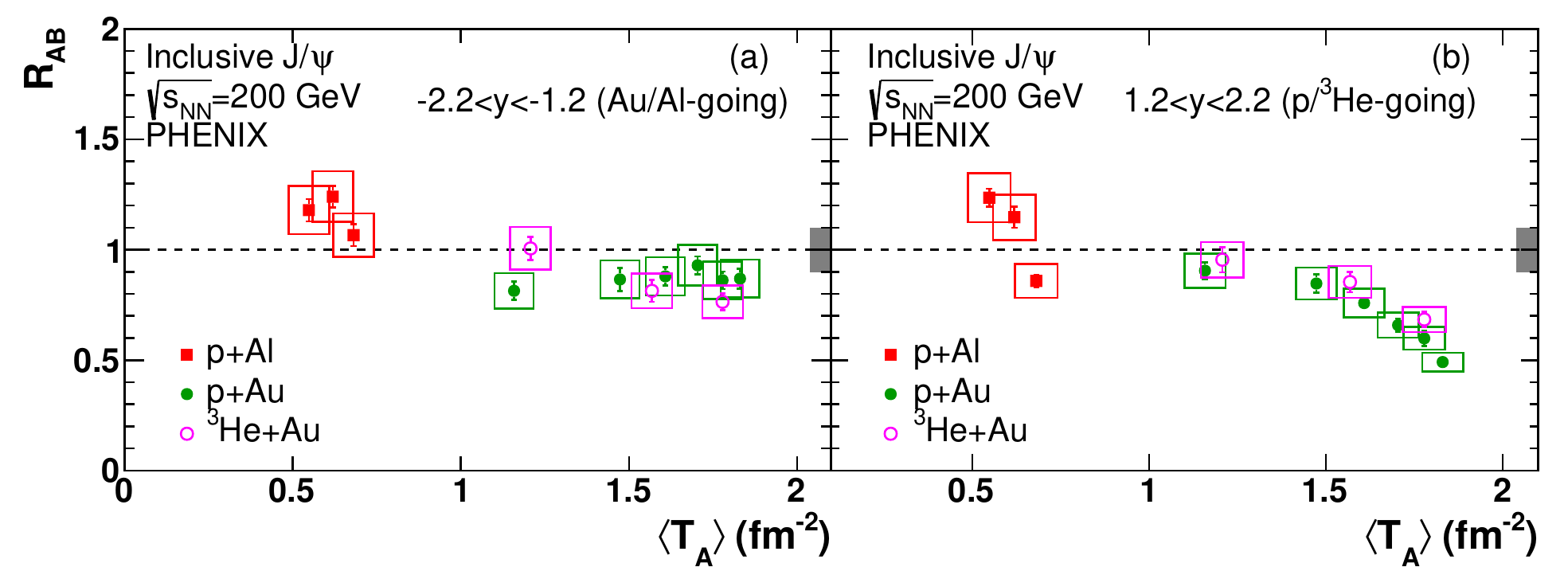}
\caption{\label{fig:rab_thick}
Nuclear modification factor of \jpsi as a function of the mean target 
thickness sampled by charmonium production in the centrality bin, for 
\pal, \pau and \heau collisions. Bars (boxes) around data points 
represent point-to-point uncorrelated (correlated) uncertainties.}
\end{figure*}

\begin{figure*}[htb]
\includegraphics[width=1.0\linewidth]{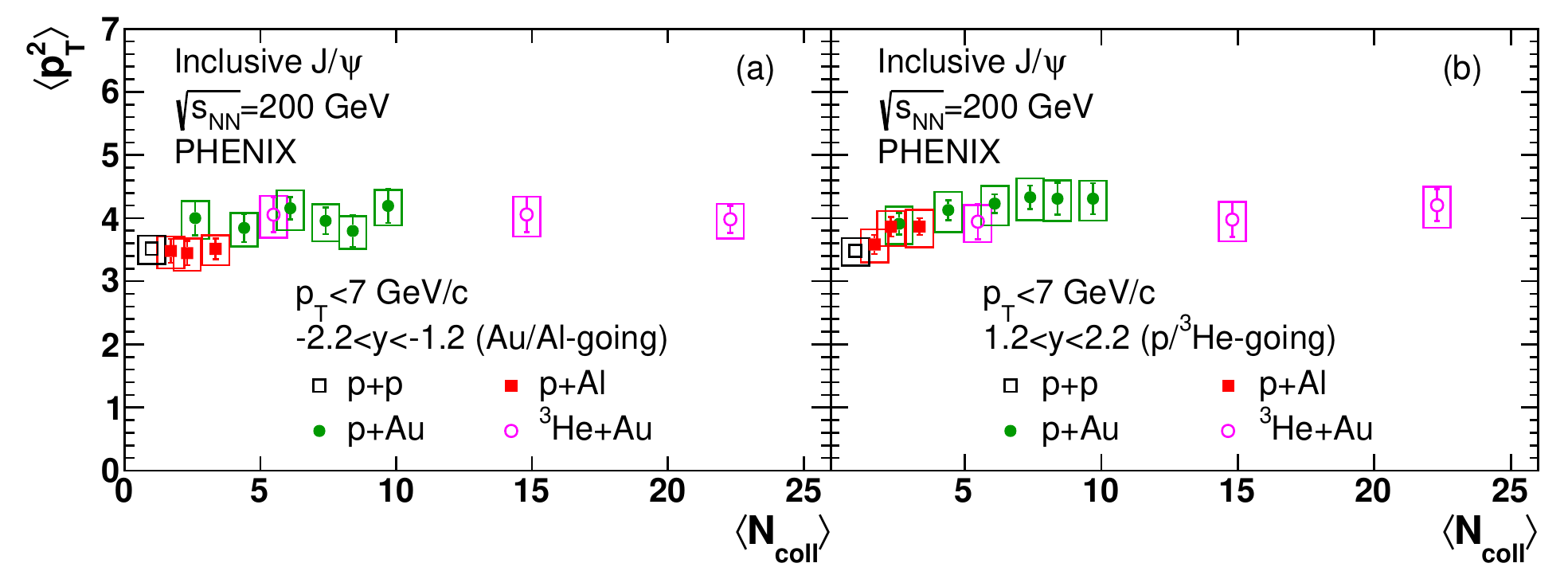}
\caption{\label{fig:ptsq_ncoll}
\meanptsq of \jpsi for $\pt<7~\mathrm{GeV}/c$ as a function of $\langle 
\Ncoll \rangle$ for \pal, \pau, and \heau collisions. Bars (boxes) 
around data points represent point-to-point uncorrelated (correlated) 
uncertainties.}
\end{figure*}


In this section, we present invariant yield, nuclear modification 
factor, and \meanptsq results at forward and backward rapidity.
There have been significant changes to the muon arm configuration and to 
the simulation framework since the \dau data set was recorded. 
Figure~\ref{fig:dndpt_pp} shows the \jpsi invariant yield as a function 
of \pt in \pp collisions at \sqstwo at forward and backward rapidity, 
where bars (boxes) represent point-to-point uncorrelated (correlated) 
uncertainties. The global systematic uncertainty is 10.1\%. The ratio of 
invariant yields between the forward and backward rapidity regions is 
presented in the bottom panel, where the systematic uncertainty due to 
the \jpsi polarization cancels in the ratio. The invariant yields at 
forward and backward rapidity are consistent within the systematic 
uncertainties, confirming that the detector efficiency is well 
understood in \pp collisions.

Plots and tables of invariant yield are presented for the other 
collision systems in the Appendix. We focus here on the nuclear 
modification factors.


Figure~\ref{fig:rab_y_mb} shows the rapidity dependence of the nuclear 
modification factor for 0\%--100\% centrality in \pal, \pau, and \heau 
collisions. The rapidity dependence of the nuclear modification for 
different centrality classes is shown for \pal in 
Fig.~\ref{fig:rab_y_pal_cent}, for \pau in 
Fig.~\ref{fig:rab_y_pau_cent}, and for \heau in 
Fig.~\ref{fig:rab_y_heau_cent}.


Figures~\ref{fig:rab_pt_mb_bwd} and~\ref{fig:rab_pt_mb_fwd} show the 
nuclear modification factor as a function of \pt for 0\%--100\% \pal, 
\pau, and \heau collisions at backward and forward rapidity. The \pt 
dependence in different centrality classes is presented for \pal in 
Fig.~\ref{fig:rab_pt_pal_cent}, for \pau in 
Figs.~\ref{fig:rab_pt_pau_cent_arm0} and~\ref{fig:rab_pt_pau_cent_arm1}, 
and for \heau in Fig.~\ref{fig:rab_pt_heau_cent}. The modification as a 
function of \pt in 0\%--20\% central collisions is compared between \pal 
and \pau in Fig.~\ref{fig:rab_pt_pal_pau_0020_fwd}.
Similar comparisons where the target is identical, but the projectile is 
different are shown for 0\%--20\% central collisions comparing \dau and \pau 
in Fig.~\ref{fig:rab_pt_dAu_pau_0020_fwd} and comparing \heau and \pau in 
Fig.~\ref{fig:rab_pt_pau_heau_0020_fwd}. 


The \pt integrated nuclear modification factor for \pal, \pau and \heau as a 
function of \meanncoll is shown at both forward and backward rapidity in 
Figs.~\ref{fig:rab_ncoll} and~\ref{fig:rab_ncoll_pAu}.  A comparison 
between \pal, \pau and \heau modifications when plotted as a function of the 
average nuclear thickness sampled by the charmonium production is presented 
in Fig.~\ref{fig:rab_thick}. Figure~\ref{fig:ptsq_ncoll} shows the mean \pt 
squared values for the three systems \pal, \pau, and \heau as a function of 
\meanncoll for $\pt<7~\mathrm{GeV}/c$ at forward and backward rapidity.


\section{Discussion}

\subsection{Rapidity dependence}

The rapidity dependence of the modification for 0\%--100\% centrality, 
seen in Fig.~\ref{fig:rab_y_mb}, shows only weak modification for \pal 
collisions. For both \pau and \heau significant suppression is seen at 
forward rapidity, with less suppression at backward rapidity. The 
modifications for \pau and \heau are very similar.

The rapidity dependence in three centrality bins for \pal collisions, 
seen in Fig.~\ref{fig:rab_y_pal_cent}, shows only weak modification in 
all centrality bins, both at forward and backward rapidity.

The \pau data presented here contain finer centrality binning for 
central collisions than was previously available from \dau. The rapidity 
dependence in six centrality bins for \pau collisions, seen in 
Fig.~\ref{fig:rab_y_pau_cent}, shows a factor of more than two 
suppression at the most forward rapidity in the 0\%--5\% centrality bin, 
and a marked increase in suppression with increasing rapidity in the 
forward direction. At backward rapidity, the modifications in all 
centrality bins show little centrality dependence, all being somewhat 
suppressed.

The rapidity dependence in three centrality bins for \heau collisions is 
shown in Fig~\ref{fig:rab_y_heau_cent}. In comparison with the \pal results 
shown in Fig.~\ref{fig:rab_y_pal_cent} for the 0\%--20\% centrality bin, 
which show little modification, the \heau results show a suppression at 
both forward and backward rapidity. The modification becomes less pronounced 
in the 20\%--40\% centrality range, and approaches unity for the most 
peripheral collisions within uncertainties.

The rapidity dependence of the 0\%--100\% centrality data is compared in 
Fig.~\ref{fig:rab_y_mb} with model calculations from 
R.~Vogt~\cite{Vogt:2015uba,Nelson:2012bc} and Shao 
\textit{et~al.}~\cite{Kusina:2017gkz,Shao:2012iz,Shao:2015vga,Lansberg:2016deg} 
showing the effect of nPDF modifications using the 
Eskola-Paakkinen-Paukkunen-Salgado (EPPS16)~\cite{Eskola:2016oht} 
next-to-leading order (NLO) and/or nuclear coordinated theoretical and 
experimental tests of quantum chromodynamics (nCTEQ15) NLO 
parameterizations~\cite{Kovarik:2015cma}. The Vogt EPPS16 NLO shadowing 
calculations in general follow the methods described in~\cite{Vogt:2015uba}, 
while the \jpsi mass and scale parameters are discussed 
in~\cite{Nelson:2012bc}.  The Shao,~\textit{et~al.} model calculations for 
\pau collisions are based on a Bayesian reweighting method which uses \jpsi 
constraints from \ppb data at the LHC~\cite{Kusina:2017gkz}. The dominant 
uncertainty in the reweighting method is the factorization scale dependence 
$\mu_F$ of the gluon modification factor R$^{Au}_{g}(x,\mu_F)$, where $\mu_F 
= \xi\mu_0$, with $\mu_0^2 = M^2 + \pt^2$ for the \jpsi transverse mass, and 
$\xi$ = 0.5, 1, 2 for the factorization scale. The reweighting however is 
not applied for lighter $^3$He and Al nuclei, with the predictions for these 
nuclei based on the original method described 
in~\cite{Shao:2012iz,Shao:2015vga,Lansberg:2016deg}. For these predictions, 
the previous PHENIX \jpsi measurement in \pp collisions~\cite{Adare:2011vq} 
is used as a baseline. The calculations were performed at all three 
factorization scales ($\mu_{0}$, $0.5~\mu_{0}$, and $2~\mu_{0}$) and provide 
two different confidence levels (68\% and 90\% CL). The uncertainty band 
shown is for the 68\% CL, and we have taken the envelope of the uncertainty 
bands from the calculations at the three scales.

In Fig.~\ref{fig:rab_y_mb}, the calculations describe the data very well at 
forward rapidity for all three collision systems, and for \pal at backward 
rapidity. For \pau and \heau at backward rapidity the calculated 
modifications are too large by roughly 40\%. However, the calculations do 
not contain effects of nuclear absorption, which is expected to be important 
at backward rapidity at \sqsntwo~\cite{McGlinchey:2012bp}, where the nuclear 
crossing time is comparable with the charmonium formation time. That is not 
expected to be the case at forward rapidity at \sqsntwo, or at the 
rapidities of interest at LHC energies. Because nuclear absorption is not 
included in the model calculations, they should be expected to overpredict 
the modification in \pau and \heau at backward rapidity.

An estimate of the effect of nuclear absorption at backward rapidity can 
be obtained from a model~\cite{Arleo:1999af} fitted to absorption cross 
sections derived from shadowing corrected data measured at a broad range 
of beam energies~\cite{McGlinchey:2012bp}. The model assumes that the 
$c\bar{c}$ pair size grows linearly with time until it reaches the size 
of a fully formed charmonium meson. Then the absorption cross section 
depends on the proper time before the pair escapes the target. The 
effect of the modification due to nuclear absorption at backward 
rapidity from this model is added to Fig.~\ref{fig:rab_y_mb}, by folding 
it into the shadowing calculation. The results indicate that the 
measured modifications are reasonably consistent with shadowing plus 
nuclear absorption.

\subsection{\pt dependence}

The \pt dependence for 0\%--100\% centrality, seen at backward rapidity 
in Fig.~\ref{fig:rab_pt_mb_bwd} and at forward rapidity in 
Fig.~\ref{fig:rab_pt_mb_fwd}, shows little modification for \pal but 
shows strong, and similar, \pt dependence for \pau and \heau. These data 
are also compared with the calculations of Shao \textit{et 
al.}~\cite{Kusina:2017gkz}. As for the rapidity dependence, the 
calculations describe the forward rapidity data well for all three 
collision systems and for the backward rapidity \pal. But the backward 
rapidity modification for \pau and \heau is overpredicted. Significant 
nuclear absorption is expected at backward rapidity and low \pt, and 
calculations that do not include it should overpredict the modification 
there.

The \pau modifications vs \pt, seen at forward rapidity in 
Fig.~\ref{fig:rab_pt_pau_cent_arm1} for all centrality bins, shows very 
strong dependence on centrality. The modification falls to 0.35 at low 
\pt for the 5\% most central collisions.  At backward rapidity, as shown 
in Fig.~\ref{fig:rab_pt_pau_cent_arm0}, the suppression is considerably 
weaker at low \pt for the most central collisions, but it changes more 
slowly with centrality. The result is that for collision centralities 
above 20\% the behavior of the modification versus \pt becomes rather 
similar at forward and backward rapidity.  
The \pt dependence of the nuclear modification factors in \pal and \heau 
collisions are shown in Figs.~\ref{fig:rab_pt_pal_cent} 
and~\ref{fig:rab_pt_heau_cent}, respectively. We see little modification 
across all three centrality ranges of \pal collisions, as was the case for 
the rapidity dependent results shown in Fig.~\ref{fig:rab_y_pal_cent}. The 
\pal nuclear modification factor for the 6--7 GeV/$c$ data point seen in 
Fig.~\ref{fig:rab_pt_mb_bwd} (a) is quite low.  However, the 6 GeV/$c$ (5--7 
GeV/$c$ bin) points for the three backward rapidity centrality bins shown in 
Fig.~\ref{fig:rab_pt_pal_cent} do not exhibit the same behavior.  We have 
therefore interpreted this last data point as being a deviation from the trend.  In 0\%--20\% \heau collisions, a suppression is observed at 
both forward and backward rapidity, and the modification becomes weaker in 
higher \pt. The modification is strongest in most central collisions, and 
the \rab approaches unity for the most peripheral collisions.  As seen with 
\pal, the last data point (5--7 GeV/$c$ bin) for \heau is also quite low.  
Likewise, we have interpreted this behavior as being a deviation from the trend, considering the measurements shown in Fig.~\ref{fig:rab_pt_mb_bwd} c) do not produce a similar effect.

The theory predictions shown in Figs.~\ref{fig:rab_pt_pau_cent_arm0} 
and~\ref{fig:rab_pt_pau_cent_arm1} are the results of adapted transport models provided by 
X. Du and R. Rapp, based on the original transport model by Zhao \& Rapp for 
$A$+$A$ collisions~\cite{Zhao:2010nk}.  The theory was extended for $d$+$A$ 
collisions~\cite{Du:2015wha} and most recently for $p$+$A$ 
collisions~\cite{Du:2018wsj}.  The transport model includes a fireball 
generated by a Monte-Carlo Glauber model~\cite{Loizides:2014vua} in addition 
to shadowing from Eskola-Paukkunen-Salgado (EPS09)~\cite{Eskola:2009uj} NLO, 
a broadening parameter~\cite{Zhao:2008pp}, and an absorption cross section 
constrained by PHENIX $d$$+$Au data~\cite{Adare:2013ezl}.  The \jpsi 
production cross section is described in~\cite{Du:2018wsj}, and charged 
particle multiplicity~\cite{Adare:2018toe}, hadronic dissociation 
rates~\cite{Du:2015wha}, and open charm production cross 
sections~\cite{Du:2018wsj} are also considered. The calculations reproduce 
the data at high \pt, but generally underpredict the suppression at low \pt 
at forward rapidity. Because the modification of \jpsi production in the 
transport model is not very strong at forward rapidity, the suppression 
there is dominated by the EPS09 shadowing contribution.

In a previous PHENIX measurement of charged particle 
multiplicity~\cite{Adare:2018toe}, it was found that twice as many particles 
are produced in 0\%--20\% central \pau collisions than in 0\%--20\% central 
\pal collisions, and the multiplicity in 0\%--20\% \heau collisions is about 
a factor of two larger than in 0\%--20\% \pau collisions. To look for 
evidence of an effect from this, 
Figs.~\ref{fig:rab_pt_pal_pau_0020_fwd},~\ref{fig:rab_pt_dAu_pau_0020_fwd}, 
and~\ref{fig:rab_pt_pau_heau_0020_fwd} show direct comparisons between the 
modifications in the 0\%--20\% centrality bin of different 
projectile~($p/d/^{3}\mathrm{He}$) and target sizes~(Al/Au). The ratio of 
nuclear modification factors is included in the bottom panel. In the 
comparisons among \pal, \pau, and \heau collisions, all systematic 
uncertainties from each collision system are included except the initial 
shape uncertainty, which cancels upon taking the ratio, and all systematic 
uncertainties stemming from the \pp system cancel.  In the comparison 
between \pau and \dau, the two systems do not share the same \pp reference, 
therefore all systematic uncertainties are included in the ratio.  Note the 
\dau data set was recorded in 2008, while the \pau data was recorded in 2015 
with a new detector.  Simulations for \dau were also performed using methods 
that differ from those used for the new small systems study.  For \pau, \dau 
and \heau comparisons, a mean value has been fitted to the ratios, and the 
result is shown on the plot together with the fit uncertainty and the 
uncertainty from the systematic errors.  The systematic uncertainty was 
determined by repeating the fit with all points moved to the upper or lower 
limits of their systematic uncertainty.

The comparison in Fig.~\ref{fig:rab_pt_pal_pau_0020_fwd} of 0\%--20\% \pal 
with 0\%--20\% \pau modifications contrasts the weak modification in central 
\pal collisions with the strong modification, particularly at forward 
rapidity, in central \pau collisions.  
Figure~\ref{fig:rab_pt_pal_pau_0020_fwd} (a) shows \pau with a nuclear 
modification factor of about 0.85 at 6 GeV/$c$ (5--7 GeV/$c$ bin). A drop in 
modification at high \pt is expected due to shadowing (and possibly also 
k$_T$ broadening). The 
comparison in Fig.~\ref{fig:rab_pt_dAu_pau_0020_fwd} of 0\%--20\% \pau with 
0\%--20\% \dau modifications highlights the similarity between the two 
systems.  A fit to the ratio of \dau to \pau at forward rapidity was found 
to be 1.13~$\pm$~0.03(stat)~$\pm$~0.13(syst) and at backward rapidity is 
0.94~$\pm$~0.02(stat)~$\pm$~0.07(syst).

In the comparison between 0\%--20\% \pau and 0\%--20\% \heau collisions 
shown in Fig.~\ref{fig:rab_pt_pau_heau_0020_fwd}, the ratio at forward 
rapidity is
\begin{equation}
\overline{R_{^{3}{\rm HeAu}}/R_{{\rm pAu}}}
= 0.96\pm0.03{\rm (stat)}\pm0.05{\rm (syst)}, \nonumber
\end{equation}
which is consistent with unity.  At backward rapidity the ratio is 
\begin{equation}
\overline{R_{^{3}{\rm HeAu}}/R_{{\rm pAu}}}
= 0.89\pm0.03{\rm (stat)}\pm0.08{\rm (syst)}.  \nonumber
\end{equation}
There may be deviations from the trend in the highest \pt bin, but large 
statistical uncertainties preclude firm conclusions.  The results are 
consistent with \jpsi production being reduced for the $^3$He projectile, 
with the backward rapidity ratio having a probability of 90\% of being less 
than one.

\subsection{\meanncoll dependence}

The \pt integrated modifications as a function of \meanncoll in each 
centrality bin are shown in Fig.~\ref{fig:rab_ncoll} for the three systems 
\pal, \pau~and \heau. No scaling with \meanncoll is expected between \pau 
and \heau, because \heau will have roughly three times as many collisions as 
\pau in the same centrality class. The \meanncoll dependence of the \pau 
modification is shown again in Fig.~\ref{fig:rab_ncoll_pAu}, where it is 
compared with the \pt integrated modification predicted by Du and Rapp. The 
theory calculation shows both the CNM baseline and the result of the 
transport calculations.  At backward rapidity some nuclear absorption is 
expected. At forward rapidity, it appears that the CNM effects are not 
strong enough to explain the data. However, the model predicts a suppression 
beyond CNM effects at backward rapidity for central collisions of 
approximately 10\%.

Modifications that are due to CNM effects (including nuclear absorption) 
would be expected to depend on the thickness of the target nucleus at 
the impact parameter of the nucleon that was involved in the hard 
process. The nuclear thickness can be written

\begin{equation}
    T_{A}(r_{T}) = \int  \rho_{A}(z, r_{T}) ~ dz,
\end{equation}
where $\rho_A(z, r_T)$ is the density distribution of nucleons in 
nucleus A taken from the Woods-Saxon distribution used in the Glauber 
model discussed in section~\ref{sec:experiment}. The parameter $z$ is 
the location in the nucleus along the beam direction, and $r_T$ is the 
transverse distance from the center of the nucleus. $T_A(r_T)$ is the 
average number of nucleons per unit area at the projectile nucleon 
impact parameter $r_T$. To get the average value of $T_A$ sampled for 
charmonium production within a given centrality bin, the values of 
$T_A(r_T)$ are weighted by the distribution of $r_T$ values within the 
centrality bin, to reflect the number of projectile nucleons having one 
or more inelastic collisions at that $r_T$, and additionally by the 
probability of a hard process at that $r_T$ -- which is proportional to 
$T_A(r_T)$.

Figure~\ref{fig:rab_thick} shows the \pal, \pau and \heau modifications 
plotted versus $\langle T_{A}\rangle$, in each centrality bin. The 
modifications seem to fall on a common curve within uncertainties, as 
would be expected if they were primarily due to CNM effects.

The \meanptsq values versus \meanncoll, shown in 
Fig.~\ref{fig:ptsq_ncoll}, fall on a common curve for all three systems. 
The \meanncoll dependence is mild, with \meanptsq increasing from 3.3 in 
\pp collisions to approximately 4.0 in \pau and \heau collisions. The 
\meanptsq is very similar between forward and backward rapidity, as was 
also observed in \dau collisions~\cite{Adare:2012qf}.

\section{Summary and conclusions}

We have presented invariant yields for inclusive \jpsi production in 
\pp, \pal, \pau and \heau collisions at \sqsntwo, and the corresponding 
nuclear modifications for \pal, \pau and \heau. The new \pau results are 
found to agree within uncertainties with the previous PHENIX \dau 
results~\cite{Adare:2010fn}.

The \pal modifications are found to be much weaker at all centralities 
than those in \pau. The 0\%--100\% centrality data for \pal are found to 
be well described in rapidity and \pt by calculations containing only 
shadowing effects from the EPPS16 NLO and nCTEQ15 NLO parameterizations, 
aside from slightly underpredicting the modification at 4--6 GeV/$c$ at 
forward rapidity.

The 0\%--100\% centrality \pau and \heau data are also compared with 
calculations based on the EPPS16 NLO and nCTEQ15 NLO shadowing 
parameterizations. At forward rapidity, the calculations describe the 
\pau and \heau modifications well in both rapidity and \pt, again with 
the exception of slightly underpredicting the modification at 4--6 
GeV/$c$ at forward rapidity. At backward rapidity, the calculations 
overpredict the modifications. We found that adding the predicted 
nuclear absorption modification taken from previous work to the backward 
rapidity \pt integrated data reduced the modifications to values 
consistent with the data.

The ratio of the \heau and \pau modifications for the 0\%--20\% 
centrality bin at forward rapidity is 

\begin{equation}
\overline{R_{^{3}{\rm HeAu}}/R_{{\rm pAu}}}
= 0.96\pm0.03{\rm (stat)}\pm0.05{\rm (syst)}, \nonumber
\end{equation}
which is smaller but consistent with unity.  At backward 
rapidity it is 

\begin{equation}
\overline{R_{^{3}{\rm HeAu}}/R_{{\rm pAu}}}
= 0.89\pm0.03{\rm (stat)}\pm0.08{\rm (syst)}. \nonumber
\end{equation}

The results are consistent with a reduction in the modification for the 
heavier projectile case. Given the systematic uncertainty, the backward 
rapidity ratio has a 90\% probability of being less than 1.0.
 
For \pau at forward rapidity, the nuclear modification vs \pt shows very 
strong centrality dependence, dropping to approximately 0.35 at low \pt 
in the most central 5\% of collisions. At backward rapidity the 
suppression is weaker for central collisions, but it changes more 
slowly. Comparison with theory calculations that include EPS09 shadowing 
and a final state transport model are able to reproduce the general 
shape of the \pt dependence at each centrality, but greatly underpredict 
the suppression at low \pt for central collisions.

The \pt integrated modification for \pau drops steeply with centrality 
at forward rapidity, reaching approximately 0.5 for the 5\% most central 
collisions. The modification at backward rapidity is found to have weak 
centrality dependence. Because nuclear absorption is evidently important 
at backward rapidity, the weak centrality dependence there is likely due 
to a trade-off between anti-shadowing and nuclear absorption. It was 
found that plotting the modification vs $\langle T_{A}\rangle$ for each 
centrality bin caused them to fall on a common line for all three 
systems, as would be expected if CNM effects dominate.


\begin{acknowledgments}

We thank the staff of the Collider-Accelerator and Physics Departments at 
Brookhaven National Laboratory and the staff of the other PHENIX 
participating institutions for their vital contributions.  We also thank 
H.-S. Shao and J.-P. Lansberg, et al., R. Vogt, X. Du, and R. Rapp for
useful discussions and providing unpublished predictions.
We acknowledge support from the Office of Nuclear Physics in the
Office of Science of the Department of Energy,
the National Science Foundation,
Abilene Christian University Research Council,
Research Foundation of SUNY, and
Dean of the College of Arts and Sciences, Vanderbilt University
(U.S.A),
Ministry of Education, Culture, Sports, Science, and Technology
and the Japan Society for the Promotion of Science (Japan),
Conselho Nacional de Desenvolvimento Cient\'{\i}fico e
Tecnol{\'o}gico and Funda\c c{\~a}o de Amparo {\`a} Pesquisa do
Estado de S{\~a}o Paulo (Brazil),
Natural Science Foundation of China (People's Republic of China),
Croatian Science Foundation and
Ministry of Science and Education (Croatia),
Ministry of Education, Youth and Sports (Czech Republic),
Centre National de la Recherche Scientifique, Commissariat
{\`a} l'{\'E}nergie Atomique, and Institut National de Physique
Nucl{\'e}aire et de Physique des Particules (France),
Bundesministerium f\"ur Bildung und Forschung, Deutscher Akademischer
Austausch Dienst, and Alexander von Humboldt Stiftung (Germany),
J. Bolyai Research Scholarship, EFOP, the New National Excellence
Program ({\'U}NKP), NKFIH, and OTKA (Hungary),
Department of Atomic Energy and Department of Science and Technology
(India),
Israel Science Foundation (Israel),
Basic Science Research and SRC(CENuM) Programs through NRF
funded by the Ministry of Education and the Ministry of
Science and ICT (Korea).
Physics Department, Lahore University of Management Sciences (Pakistan),
Ministry of Education and Science, Russian Academy of Sciences,
Federal Agency of Atomic Energy (Russia),
VR and Wallenberg Foundation (Sweden),
the U.S. Civilian Research and Development Foundation for the
Independent States of the Former Soviet Union,
the Hungarian American Enterprise Scholarship Fund,
the US-Hungarian Fulbright Foundation,
and the US-Israel Binational Science Foundation.

\end{acknowledgments}

\section*{appendix}

\begin{figure*}[tbh]
\includegraphics[width=0.67\linewidth]{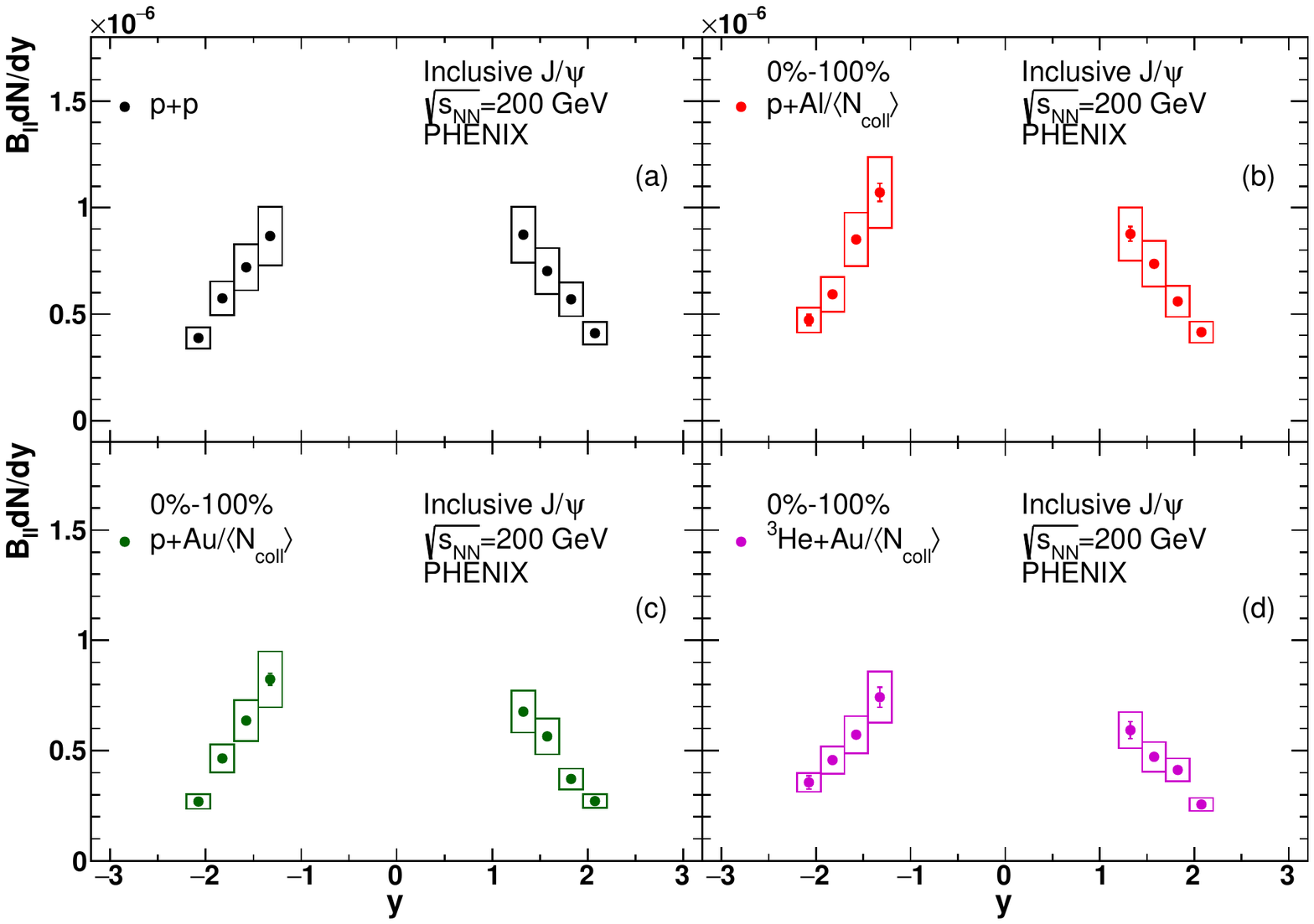}
\caption{\label{fig:dndy_mb}
\jpsi invariant yield as a function of $y$ in MB \pp, \pal, \pau, and 
\heau collisions. Bars (boxes) around data points represents 
point-to-point uncorrelated (correlated) uncertainties.  There is also a 
global uncertainty of 10.1\%, 11.5\%, 12.1\% and 12.2\% corresponding to 
\pp, \pal, \pau and \heau yields.}
\end{figure*}

\begin{figure*}[hb]
\begin{minipage}{0.98\linewidth}
\includegraphics[width=1.0\linewidth]{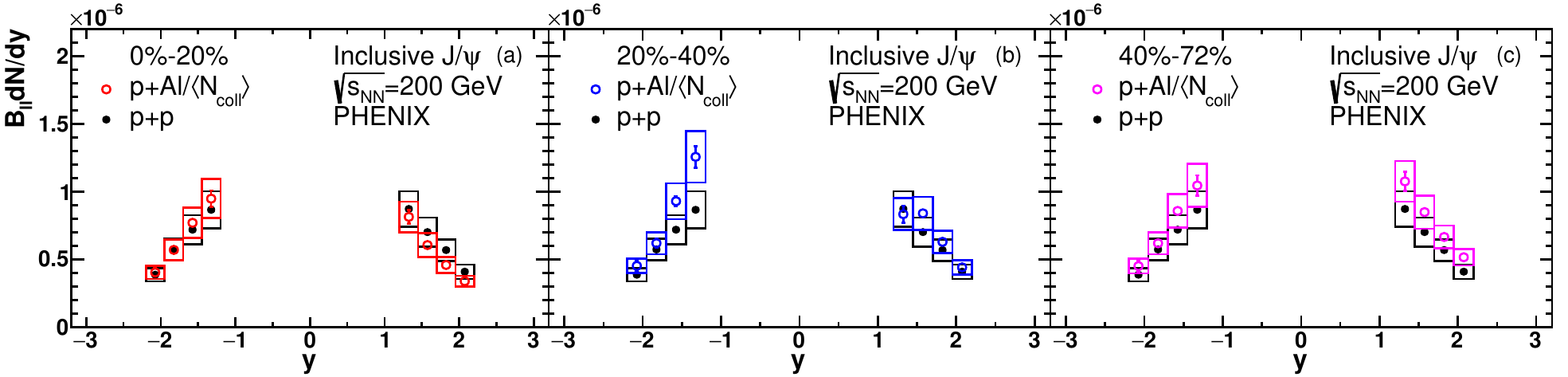}
\caption{\label{fig:dndy_pal}
\jpsi invariant yield as a function of $y$ in various centrality bins of 
\pal collisions. Bars (boxes) around data points represents 
point-to-point uncorrelated (correlated) uncertainties.  There is also a 
global uncertainty of 13.6\%, 12.2\%, and 12.3\% corresponding to 
0\%--20\%, 20\%--40\% and 40\%--72\% centrality.}
\end{minipage}
\begin{minipage}{0.98\linewidth}
\includegraphics[width=1.0\linewidth]{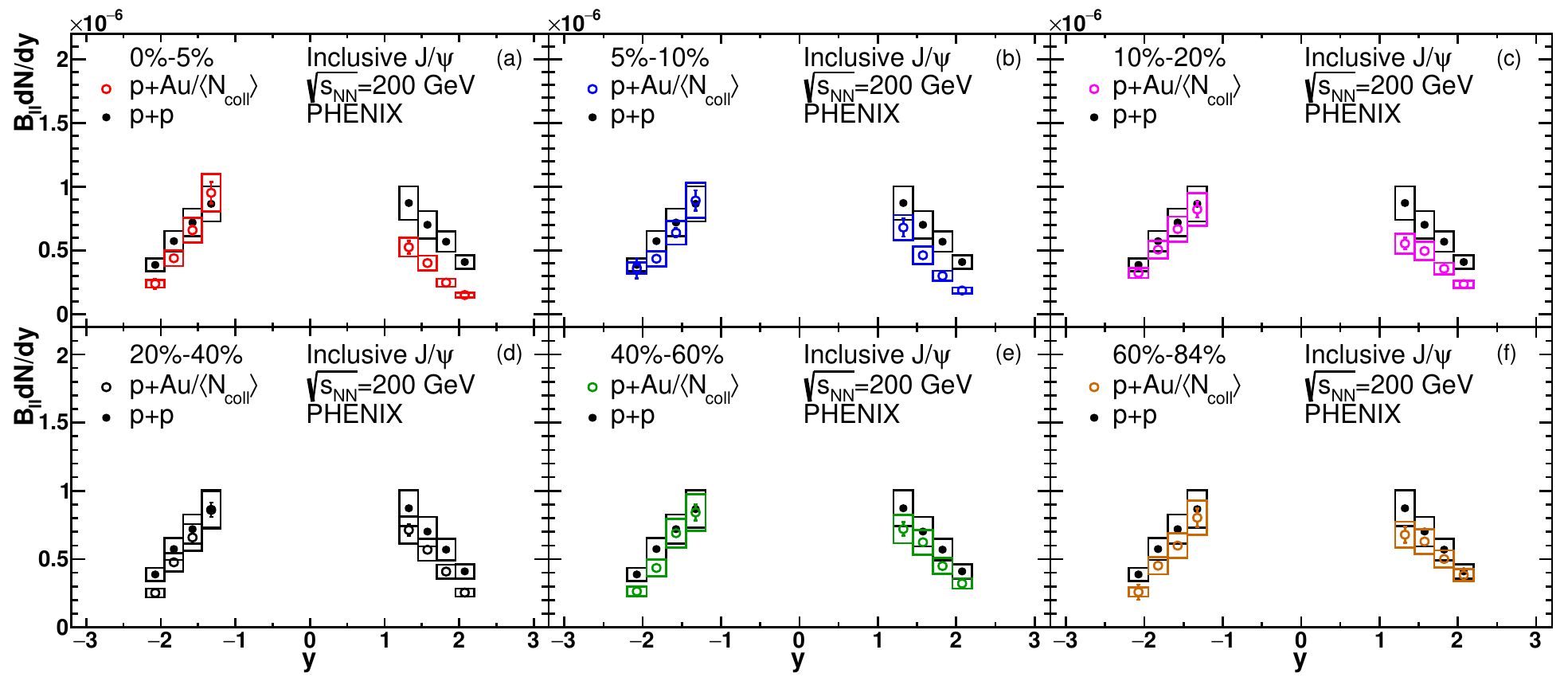}
\caption{\label{fig:dndy_pau}
\jpsi invariant yield as a function of $y$ in various centrality bins of 
\pau collisions. Bars (boxes) around data points represents 
point-to-point uncorrelated (correlated) uncertainties.  There is also a 
global uncertainty of 11.9\%, 11.8\%, 12.2\%, 12.1\%, 12.2\% and 14.0\% 
corresponding to 0\%--5\%, 5\%--10\%, 10\%--20\%, 20\%--40\%, 
40\%--60\%, and 60\%--84\% centrality.}
\end{minipage}
\begin{minipage}{0.98\linewidth}
\includegraphics[width=1.0\linewidth]{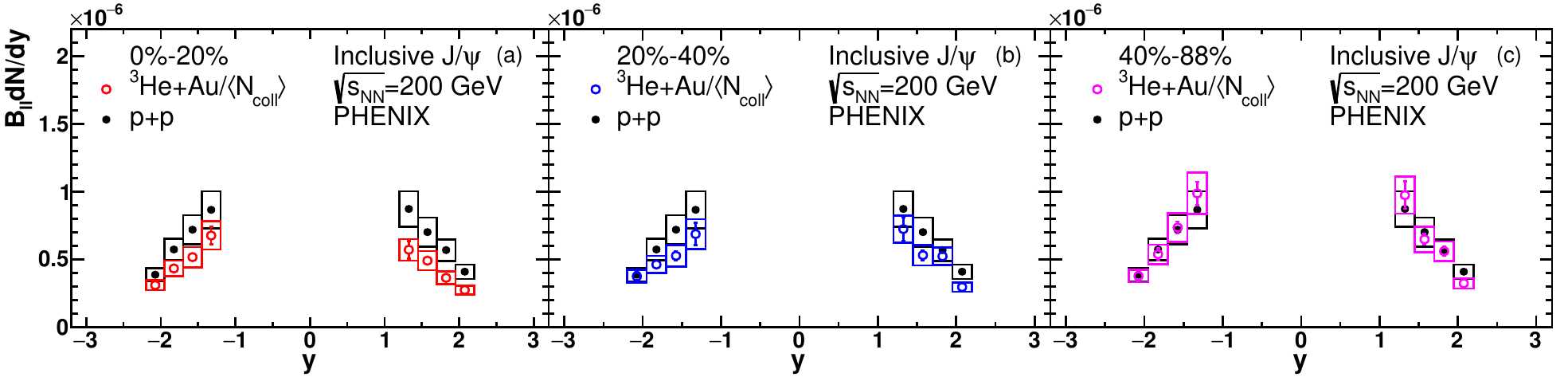}
\caption{\label{fig:dndy_heau}
\jpsi invariant yield as a function of $y$ in various centrality bins of 
\heau collisions. Bars (boxes) around data points represents point-to-point 
uncorrelated (correlated) uncertainties. There is also a global uncertainty 
of 12.7\%, 12.6\%, and 13.4\% corresponding to 0\%--20\%, 20\%--40\%, and 
40\%--88\% centrality.}
\end{minipage}
\end{figure*}
\begin{figure*}[htb]
\begin{minipage}{0.98\linewidth}
\includegraphics[width=1.0\linewidth]{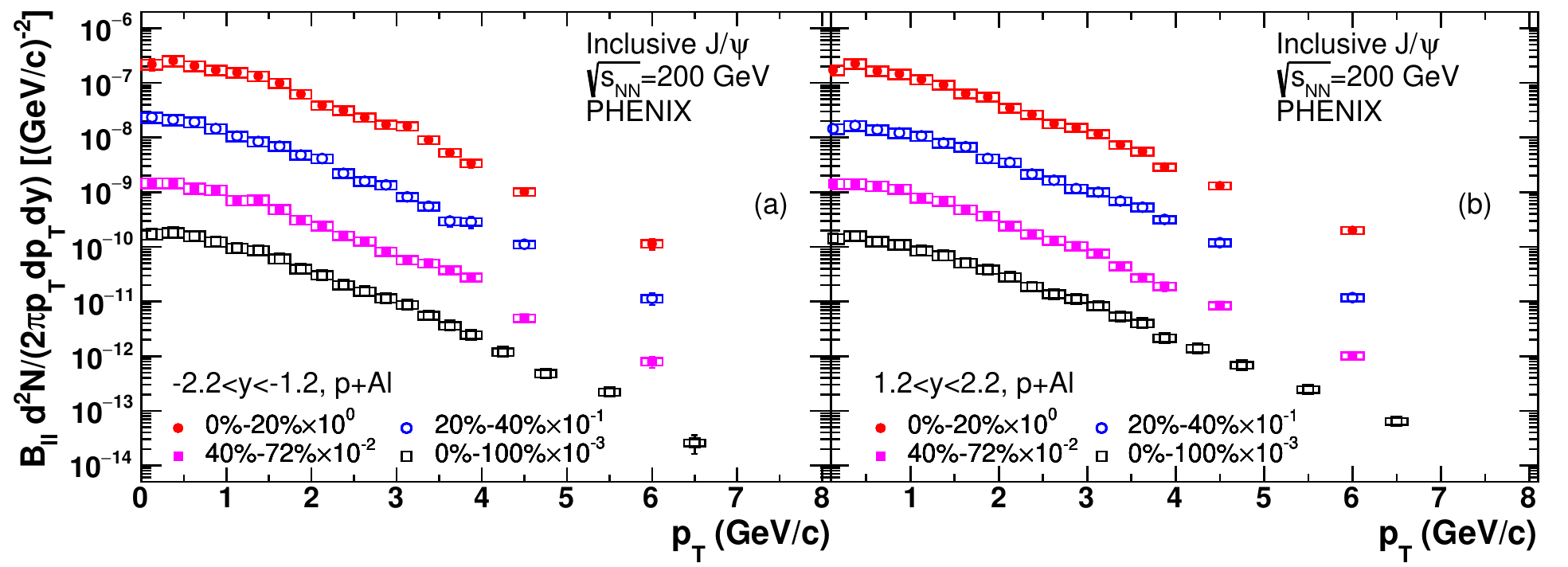}
\caption{\label{fig:dndpt_pal}
\jpsi invariant yield as a function of \pt in various centrality bins of 
\pal collisions, and the yields in each centrality bin are scaled for 
better visibility. Bars (boxes) around data points represents 
point-to-point uncorrelated (correlated) uncertainties.  There is also a 
global uncertainty of 10.2\%, 10.3\%, 10.9\%, and 10.4\% corresponding to 
0\%--20\%, 20\%--40\%, 40\%--72\%, and 0\%--100\% centrality.}
\end{minipage}
\begin{minipage}{0.98\linewidth}
\includegraphics[width=1.0\linewidth]{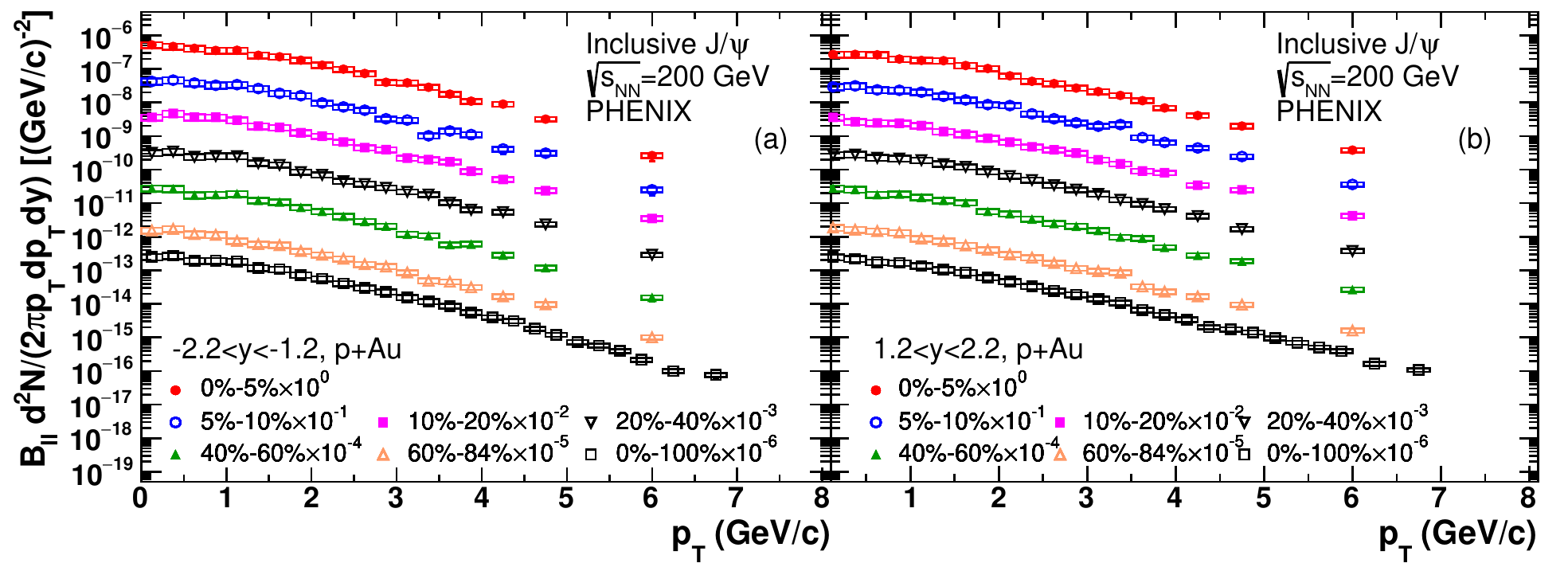}
\caption{\label{fig:dndpt_pau}
\jpsi invariant yield as a function of \pt in various centrality bins of 
\pau collisions, and the yields in each centrality bin are scaled for 
better visibility. Bars (boxes) around data points represents 
point-to-point uncorrelated (correlated) uncertainties.  There is also a 
global uncertainty of 11.8\% for 60\%--84\% centrality and 10.2\% for 
all remaining centralities.}
\end{minipage}
\begin{minipage}{0.98\linewidth}
\includegraphics[width=1.0\linewidth]{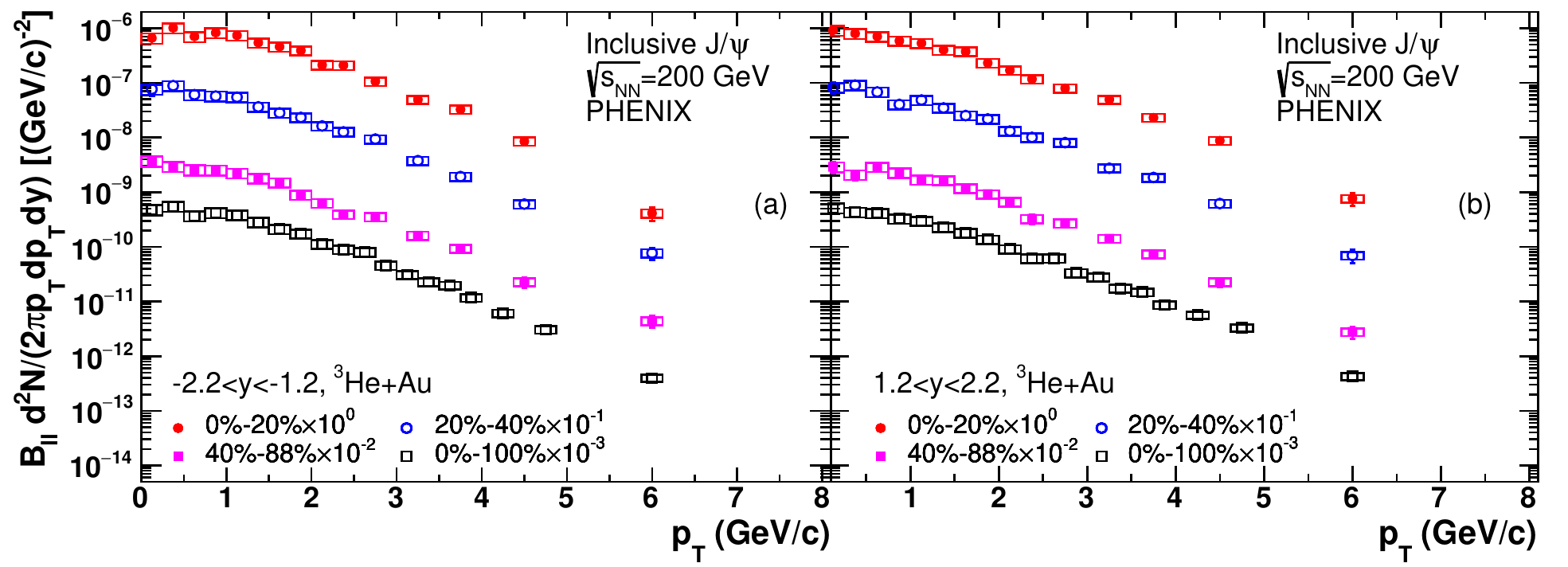}
\caption{\label{fig:dndpt_heau}
\jpsi invariant yield as a function of \pt in various centrality bins of 
\heau collisions, and the yields in each centrality bin are scaled for 
better visibility. Bars (boxes) around data points represents 
point-to-point uncorrelated (correlated) uncertainties.  There is also a 
global uncertainty of 10.7\% for 40\%--88\% centrality and 10.2\% for 
all remaining centralities.}
\end{minipage}
\end{figure*}

The invariant yields for all data sets are presented in this appendix. 
Figure~\ref{fig:dndy_mb} shows inclusive \jpsi invariant yield as a 
function of rapidity in MB \pp, \pal, \pau, and \heau collisions, and 
the invariant yields in \pal, \pau, and \heau collisions are scaled with 
\meanncoll to compare with the invariant yield in \pp collisions. In 
this and the following figures showing results of invariant yield 
measurement, the bars (boxes) around data points represent point-to-point 
uncorrelated (correlated) uncertainties. 
Figures.~\ref{fig:dndy_pal},~\ref{fig:dndy_pau}, and~\ref{fig:dndy_heau} 
show inclusive \jpsi invariant yield as a function of rapidity in 
different centrality of \pal, \pau, and \heau collisions, respectively. 
Invariant yields in \pal, \pau, and \heau collisions are scaled with 
\meanncoll, and the \pp result is also presented in each panel. 
Figures~\ref{fig:dndpt_pal},~\ref{fig:dndpt_pau}, 
and~\ref{fig:dndpt_heau} show inclusive \jpsi invariant yield as a 
function of \pt in different centrality of \pal, \pau, and \heau 
collisions, respectively. 

At $\pt>2.5~\mathrm{GeV}/c$, \pt binning was changed for different data sets 
depending on statistics as described in Table~\ref{tab:ptbin}. When 
calculating the nuclear modification factor for \pt bins of different 
$\Delta\pt$ from the \pp data, additional fits to the \pp data were 
performed to match the \pt binning of the $p/^{3}\mathrm{He}$+$A$ data.

\begin{table*}[tbh]
\caption{\label{tab:ptbin}
\pt bins in different data sets and centrality bins. All values are in $\mathrm{GeV}/c$.
}
\begin{ruledtabular} \begin{tabular}{ccccccc}
\pp & \pal & \pal & \pau & \pau & \heau & \heau\\
 & 0\%--100\% & Centrality & 0\%--100\% & Centrality & 0\%--100\% & Centrality\\\hline
0.00--0.25 & 0.00--0.25 & 0.00--0.25 & 0.00--0.25 & 0.00--0.25 & 0.00--0.25 & 0.00--0.25\\
0.25--0.50 & 0.25--0.50 & 0.25--0.50 & 0.25--0.50 & 0.25--0.50 & 0.25--0.50 & 0.25--0.50\\
0.50--0.75 & 0.50--0.75 & 0.50--0.75 & 0.50--0.75 & 0.50--0.75 & 0.50--0.75 & 0.50--0.75\\
0.75--1.00 & 0.75--1.00 & 0.75--1.00 & 0.75--1.00 & 0.75--1.00 & 0.75--1.00 & 0.75--1.00\\
1.00--1.25 & 1.00--1.25 & 1.00--1.25 & 1.00--1.25 & 1.00--1.25 & 1.00--1.25 & 1.00--1.25\\
1.25--1.50 & 1.25--1.50 & 1.25--1.50 & 1.25--1.50 & 1.25--1.50 & 1.25--1.50 & 1.25--1.50\\
1.50--1.75 & 1.50--1.75 & 1.50--1.75 & 1.50--1.75 & 1.50--1.75 & 1.50--1.75 & 1.50--1.75\\
1.75--2.00 & 1.75--2.00 & 1.75--2.00 & 1.75--2.00 & 1.75--2.00 & 1.75--2.00 & 1.75--2.00\\
2.00--2.25 & 2.00--2.25 & 2.00--2.25 & 2.00--2.25 & 2.00--2.25 & 2.00--2.25 & 2.00--2.25\\
2.25--2.50 & 2.25--2.50 & 2.25--2.50 & 2.25--2.50 & 2.25--2.50 & 2.25--2.50 & 2.25--2.50\\
2.50--2.75 & 2.50--2.75 & 2.50--2.75 & 2.50--2.75 & 2.50--2.75 & 2.50--2.75 & 2.50--3.00\\
2.75--3.00 & 2.75--3.00 & 2.75--3.00 & 2.75--3.00 & 2.75--3.00 & 2.75--3.00 & 3.00--3.50\\
3.00--3.25 & 3.00--3.25 & 3.00--3.25 & 3.00--3.25 & 3.00--3.25 & 3.00--3.25 & 3.50--4.00\\
3.25--3.50 & 3.25--3.50 & 3.25--3.50 & 3.25--3.50 & 3.25--3.50 & 3.25--3.50 & 4.00--5.00\\
3.50--3.75 & 3.50--3.75 & 3.50--3.75 & 3.50--3.75 & 3.50--3.75 & 3.50--3.75 & 5.00--7.00\\
3.75--4.00 & 3.75--4.00 & 3.75--4.00 & 3.75--4.00 & 3.75--4.00 & 3.75--4.00 \\
4.00--4.25 & 4.00--4.50 & 4.00--5.00 & 4.00--4.25 & 4.00--4.50 & 4.00--4.50\\
4.25--4.50 & 4.50--5.00 & 5.00--7.00 & 4.25--4.50 & 4.50--5.00 & 4.50--5.00\\
4.50--4.75 & 5.00--6.00 & & 4.50--4.75 & 5.00--7.00 & 5.00--7.00\\
4.75--5.00 & 6.00--7.00 & & 4.75--5.00\\
5.00--5.25 & & & 5.00--5.25\\
5.25--5.50 & & & 5.25--5.50\\
5.50--5.75 & & & 5.50--5.75\\
5.75--6.00 & & & 5.75--6.00\\
6.00--6.50 & & & 6.00--6.50\\
6.50--7.00 & & & 6.50--7.00\\
\end{tabular} \end{ruledtabular}
\end{table*}

\clearpage

 
%
 
\end{document}